# Trust in AI: Progress, Challenges, and Future Directions

**Saleh Afroogh[1]\*  Ali Akbari[2]  Evan Malone [3]  Mohammadali Kargar [4]  Hananeh Alambeigi [5]**

*1. Department of Philosophy, The University of Texas at Austin, Austin, NY 12203, USA. saleh.afroogh@utexas.edu*

*2. Stanford University School of Medicine, 450 Serra Mall, Stanford, CA 94305. akbaria@stanford.edu.*

*3. Department of Philosophy, Lone Star College in Houston, TX. evan.c.malone@lonestar.edu*

*4. Department of Mechanical Engineering, Texas A&M University, College Station, TX, USA. mohammadkrg@tamu.edu*

*5.Department of Industrial and Systems Engineering, Texas A&M University, College Station, TX, USA hana.alambeigi@tamu.edu*

\* Correspondence: saleh.afroogh@utexas.edu

The increasing use of artificial intelligence (AI) systems in our daily life through various applications, services, and products explains the significance of trust/distrust in AI from a user perspective. AI-driven systems (as opposed to other technologies) have ubiquitously diffused in our life not only as some beneficial tools to be used by human agents but also are going to be substitutive agents on our behalf, or manipulative minds that would influence human thought, decision, and agency. Trust/distrust in AI plays the role of a regulator and could significantly control the level of this diffusion, as trust can increase, and distrust may reduce the rate of adoption of AI. Recently, varieties of studies have paid attention to the variant dimension of trust/distrust in AI, and its relevant considerations. In this systematic literature review, after conceptualization of trust in the current AI literature review, we will investigate trust in different types of human-Machine interaction, and its impact on technology acceptance in different domains. In addition to that, we propose a taxonomy of technical (i.e., safety, accuracy, robustness) and non-technical axiological (i.e., ethical, legal, and mixed) trustworthiness metrics, and some trustworthy measurements. Moreover, we examine some major trust-breakers in AI (e.g., autonomy and dignity threat), and trust makers; and propose some future directions and probable solutions for the transition to a trustworthy AI.

# 1. INTRODUCTION

A person's trust in someone or something can determine their behavior, interaction, and acceptance [1]. In fact, trust is a crucial factor in accepting and adopting technology in real life. Artificial intelligence (AI), defined as the ability of a machine or a system to perform human-like tasks [2], has widely diffused in our everyday daily life through various applications, services, and products. AI is an integral part of modern life, playing an increasingly important role in our daily life [1]. AI has achieved significant progress in outperforming conventional solutions in many areas, including health [2]–[4], autonomous transportation [5]–[7], military [8], [9], data security [10], [11], entertainment [12]–[15] etc. This has led to the rapid increase of AI-based methods in these areas.

Trust in AI can significantly control the level of this diffusion as distrust may reduce the chance of adoption of AI. Trust in AI can be viewed as "the willingness of people to accept AI and believe in the suggestions, decisions made by the system, share tasks, contribute information and provide support to such technology" [1]. AI can be developed and adopted only if it satisfies the stakeholders' and users' expectations and needs, and that is how the role of trust becomes essential. In general terms, trust is built when the trustor can anticipate the trustee's behavior to know if it matches its desires [18]. Therefore, individuals, organizations, and societies will only ever be able to realize the full potential of AI if trust can be established in its development, deployment, and use [19]. Therefore, it is vital to understand the definition, scope, and role of trust in AI technology and determine its influential factors and unique application-dependent requirements.

Trust in AI is not just a non-technical ethical consideration [20]. Instead, it also includes various domains, including AI performance, transparency and explainability, and compliance with legal and technical regulations. AI is different from other automated systems in the sense that it can learn, and it can behave proactively, unexpectedly, and incomprehensibly for humans [21]. Overall, influential factors of trust in technology could be divided into human-based, context-based, and technology-based factors. No matter what technology the trustee is, the impacts of human-based and context-based factors are more or less similar. For instance, a person with a high-trusting stance would be more likely to accept and depend on new technologies [1]. However, the technology-based factors of AI that affect trust are unique and usually more challenging than other technologies, even compared to rule-based automation. That is because, in AI, the system can make new decisions based on training data. Therefore, parameters such as accuracy, reliability, transparency, and explainability of the decision become extremely important to determine the level of trustworthiness of AI.

Recently, many researchers have tried to identify reasons for distrust in AI and improve trust by different means since distrust has hindered the successful adoption of AI technology in various domains. For instance, despite AI's considerable potential in the manufacturing industry, its application still faces the challenge of insufficient trust due to the black-box nature of AI that introduces difficulties for ordinary users to understand it [22]. Medical imaging is another domain that can significantly benefit from AI technology. Still, these technologies have not been widely adopted in this area due to lack of trust by medical practitioners, healthcare stakeholders, and patients, in addition to regulatory, medicolegal, or ethical issues [23]. Similarly, risk-averseness and lack of trust have limited the application and adoption of AI technology in many other domains such as autonomous vehicles [24], customer service chatbots [25], personal assistants [26], finance [27], [28], depression treatment [29], robotics [30], and IoT [31]. Accordingly, trust in AI functions as a driver in AI usage, and distrust is considered a barrier to the development and application of AI systems, and it would negatively affect the stakeholder's perspective toward AI systems in different contexts.

Different dimensions and impacts of trust/distrust in AI are mentioned and discussed in a variety of studies, reports, and case studies in different domains. The current body of knowledge, however, lacks a systematic review of the different dimensions and varying considerations of conceptualization of trust/distrust in AI, and a discussion of the relationships and possible resolutions of these considerations. Therefore, in this study, we conduct a systematic literature review to 1) reveal the different conceptions and theories of trust/distrust in AI, as well as its different types, models, and relevant impacts; 2) discuss the two major classes of technical and axiological trustworthy metrics, relevant frameworks and measurements, as well as distrust origins and motivations, such as autonomy and dignity threat; 3) provide solutions for some problems and considerations that accelerate the transition to a trustworthy and responsible AI.

Our discussion proceeds as follows (see, Table 1): *2. Methodology,* describes the methods used in systematic reviews of studies related to trust in AI. *3. Findings,* presents the findings and results related to the key values and major cords and how they are discussed in the literature. It includes the following subsections: 3.1.Different types/models of trust in AI, 3.2.Trustworthy AI and its metrics: Trustworthy AI, Distrust in AI and Scary AI, 3.4. Trust makers: building/increasing trust in AI. *4. Discussion,* analytically discusses the major codes and key values and considerations related to trust in AI, as well as the address of the practical value conflicts and the probable trade-off between the key values and considerations. It includes 12 subsections. Concluding Remarks and Future Directions for trust research in AI are also discussed in section 5.

**Table 1: A road map of this study**

| Section N. | Section Title | Subsection themes |
|:---:|:---|:---|
| **1** | **Introduction** | |
| **2** | **Methodology** | |
| **3** | **Findings** | |
| | 3.1. Different types/models of trust in AI (Human-Machine interaction) | 3.1.1. Theories and definitions of trust in AI |
| | | 3.1.2. Trust in types of human-Machine interaction |
| | | 3.1.3. Impact of trust/distrust on AI technology acceptance in different domains |
| | 3.2. Trustworthy AI and its metrics: Trustworthy AI (i.e., technical, and non-technical metrics: legal, ethical, mixed) | 3.2.1. Trust & explainability / transparency / interpretability |
| | | 3.2.2. Trust & empathy in AI |
| | | 3.2.3. Trust and privacy |
| | | 3.2.4. Trust and fairness in AI |
| | | 3.2.5. Trust and accountability in AI |
| | | 3.2.6. Trust and technical metrics (safety, accuracy, robustness,) |
| | | 3.2.7. Evaluating and measuring/ trustworthiness certificate in AI |
| | | 3.2.8. Trustworthy AI Frameworks |
| | 3.3. Distrust in AI and Scary AI | 3.3.1. Distrust makers in AI systems |
| | | 3.3.2. Surveillance, and manipulation |
| | | 3.3.3. Human autonomy/dignity threat |
| | | 3.3.4. Distrust and unpredictable futures |
| | | 3.3.5. Challenges and barriers to breaking distrust |
| | 3.4. Trust makers: building/increasing trust in AI | 3.4.1. Factors that affect trust |
| | | 3.4.2. Methods of Building trust in AI |
| | | 3.4.3. Case studies and items effects on building trust |
| **4** | **Discussion** | 4.1. Interaction of technical and non-technical factors of trust and trustworthiness in AI |
| | | 4.2. Non-interchangeability of interpretability, explainability and transparency, and their classification |
| | | 4.3. Trust as a two-way street |
| | | 4.4. Distinction between empathy in human's trust in AI and empathy in AI's trust in human agents |
| | | 4.5. Tradeoff between empathy and privacy |
| | | 4.6. The subjectivity of trust in AI vs. the objectivity of reliable AI |
| | | 4.7. AI privacy and human agent privacy |
| | | 4.8. The developmental problem of 'right to explanation' |
| | | 4.9. Development of direct AI accountability |
| | | 4.10. Challenges of measuring trust and trustworthiness in AI |
| | | 4.11. Trust equity problem in AI |
| | | 4.12. Impossibility of Interpersonal trust in AI systems |
| **5** | **Concluding Remarks and Future Directions** | |

## 2. METHODOLOGY

We conducted an inclusive and systematic review of academic papers, reports, case studies, and trust frameworks in AI, written in English. Given that there is not a specific database on trust in AI in particular, we used the Preferred Reporting Items for Systematic Reviews and Meta-Analyses (PRISMA) framework to develop a protocol in this review (Figure 1).

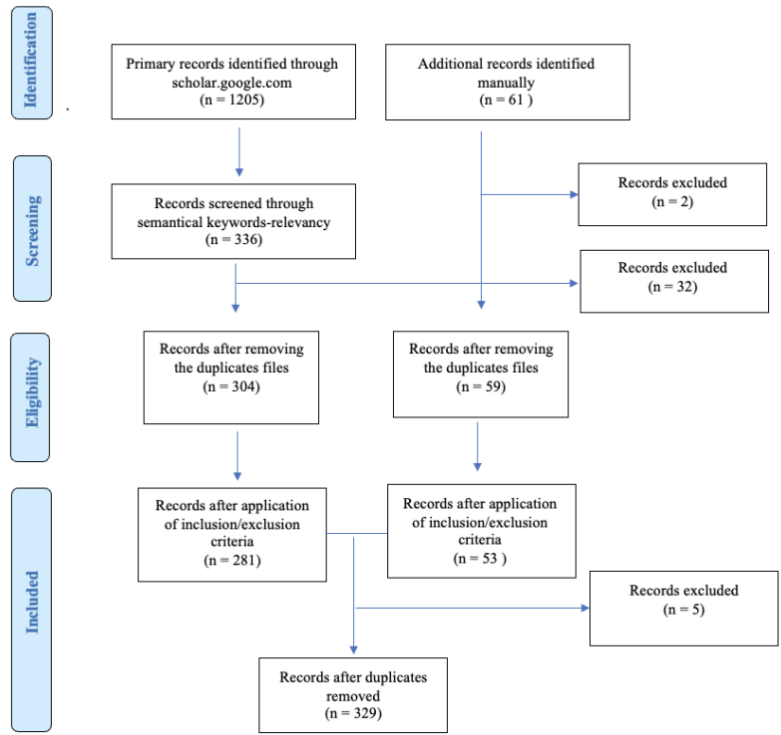

**Figure 1: Developed PRISMA flow diagram for review of trust in AI.**

In order to conduct a comprehensive review of the relevant studies, we followed two approaches. First, we manually searched for the most related papers on trust in AI: 19 papers were identified through the online search after removal of duplicate files. Secondly, we fulfilled a keyword-based search (using the http://scholar.google.com search engine) to collect all relevant papers on the topic. This search was accomplished using the following keyword phrases: (1) "trust + AI" which provided 19 relevant result pages of Google Scholar, (2) "trust+Artificial+Intelligence" for which the first five result pages were reviewed, (3) "trustworthy + AI," for which the first 15 result pages were reviewed; and (4) "trustworthy+Artificial+Intelligence ," for which the first 13 result pages of Google Scholar were reviewed. Moreover, the following keywords "Trust + explainability/transparency/ interpretability/empathy/privacy/fairness/accountability/safety/accuracy/robustness + AI /Artificial+Intelligence" and "distrust + AI/ Artificial + Intelligence" were included because of their central role in the research as the major known (based on a preliminary review) considerations of trust in AI. Additionally, the search was suspended within results for each search term due to limited appearances of new relevant papers on the following pages.

The results of the search were 336 relevant papers (which were selected based on the semantical keywords relevancy), out of 1205 (which appeared on the result pages). Afterward, the duplicated papers were eliminated from the analysis. We selected the 329 target papers for this systematic review based on the following two inclusion/exclusion criteria. First, articles that were published in academic journals were included. Second, the dominant topic of the papers (or a significant part of it) was trust in AI. To this end, the papers' main sections were reviewed to understand their dominant topic rather than only relying on the title and papers' keywords.

## 3. FINDINGS

The qualitative analysis on the selected papers was performed by four researchers who critically read the papers and who developed the eight major key codes as the building blocks of the categorization of the review result in the next step of this research (Table 2).

### Table 2: Major and minor codes included in the reviewed papers

| Major ethical codes | N. of reviewed papers | Minor ethical codes |
|---|---|---|
| Theories and definitions of trust in AI | 16 | Interaction between human and AI, directional transaction, vulnerability acceptance, facilitating collaboration, risk and uncertainty, model's "correctness", confident decision, integrity and ability, user's perception of an AI system's ability, aesthetic of a user interface, behavior and risk anticipation, reliability, perceived trustworthiness, predictive power, over-trust, beneficence, non-maleficence, autonomy, justice, and explicability |
| **Trust in types of human-machine interaction** | 25 | human-machine interactions, machine-human interactions, machine-machine interactions, machines as the host of AI, Robo-advisors, autonomous vehicles, Adversarial attacks, unreliable sources, smart contracts, self-imposed standards, certification, corporate guidelines, governmental regulations |
| **Impact of trust/distrust on AI technology acceptance in different domains** | 31 | economic output, electronic markets, acceptance of AI by physicians, Reduce the wait time in healthcare, algorithmic investment advice, AI-based personal assistants, chatbots, [AI] deception, racist and genocidal ideologies in [AI developers], managers' endorsement, cognitive trust, emotional trust |
| **Trust & explainability/ transparency/interpretability in AI** | 53 | complex opaque concepts, deep neural networks, opaque nature of complex AI algorithms, AI-based decisions, transparency vs. explainability, transparency against overtrusting AI, levels of transparency, dynamic process [of trust building], transparency criteria, [right of rejecting] automated processing, interpretability vs. explainability, [AI's] decision's rationale, human-interpretable, model's inner machinery, pre-model interpretability, intrinsic interpretability, post-hoc interpretability |
| **Trust & empathy in AI** | 12 | subjective process, deep understanding of other people's feelings, non-judgmental, ability to simulate, cognitive ability, accurate inference, empathic accuracy, affective or emotional ability, supportive, benevolent, and compassionate response, other's feelings and thoughts, efficient communication, social bonding, social interactions, people's mental states, user's expectations, agent's credibility and trust, empathic cultural-aware agents, social values and norms, behavioral and motivational levels, observation and detection of social signals, empathic action and interaction, stakeholders' viewpoints, similar tastes and ratings |
| **Trust and privacy** | 20 | personal information, detailed information, customer privacy empowerment, privacy-by-design, data minimization, controllability, transparency, easy-to-use privacy function, data confidentiality, different levels of privacy, a pessimism problem, |
| **Trust and fairness in AI** | 9 | Algorithmic bias, discrimination, perceived fairness, induced fairness, equal treatment, regulations, Algorithmic unfairness, minority and marginalized groups, User biases, perceptions of harm and injustice, reported wrongdoing, rules and regulations Implementation |
| **Trust and accountability in AI** | 5 | legal framework, public trust in AI, reliability of models, minimum standards, Transparent explanations, empathy, privacy concerns |
| **Trust and technical metrics (safety, accuracy, robustness,) in AI** | 22 | vulnerability of the user, technical elements of trust, Reliability, security, lineage, system accuracy, weight of advice, high-stakes decisions, calibrating the trust in AI systems, distrust AI, over-trust AI, zero-touch security, diversity, segregating malicious nodes, cybersecurity, Bayesian-based trust model, human multi-robot team, biosignals |
| **Evaluating and measuring/ trustworthiness certificate in AI** | 17 | psycho-physiological approaches, empirical approaches, theoretical methods, accuracy and safety guidelines, qualitative evaluations, experimental constraints, trust-theoretical model, |
| | | transparent AI systems, vulnerability and risk assessments, physiological model, multi-dimensional metrics, ndividual and team performance scores, situation awareness, philosophical evaluation of trust, complexity and inexplicability, algorithmic auditing, customization of AI certification, resilience, agility, satisfaction, efficiency, data protection, predictability, believability |
| **Trust Frameworks in AI** | 22 | ethical AI systems, flexibility, accurate incorporation of the data, privacy protection, ethics by design, ethics in design, ethics for design, algorithmic bias, worldwide health, fairness, accountability, transparency, behavioral patterns, invitation of trust, culture factors, explainability, multi-level framework, robot autonomy, sociological framework, unwarranted varieties of trust, warranted trustworthy AI, Human agency, technical robustness and safety, data governance, privacy, diversity, societal and environmental wellbeing, trust measurement, justice and fairness, non-maleficence |
| **Distrust in AI** | 26 | Scary AI, faster than human beings, malevolent artificial intelligence, alignment problem, surveillance, privacy, hackability, loss of human control, [AI's] uses in war, applications in healthcare, potential consequences of AI for the economy, Distrust makers in AI, surveillance & manipulation, human autonomy, dignity thread, and unpredictable futures, distrust breaker, optimal level of transparency, bias propagation |
| **Trust makers: Building/increasing trust in AI** | 33 | technical and axiological factors, AI personality, anthropomorphism, reputation, transparency, team-related factors, context-related factors, individual-related factors, human traits, actual capabilities of the AI, human agency and oversight, accountability, marketing, over-trust, technical method for building trust, global and local explainability, local justifications, interactive visualization, sharing transparency, standard or technical regulation, non-expert end-users, experts' endorsement, AI risk-mitigating practices, certification/accreditation, performance metrics, responsibility of AI system, interpretations of these standards, Value-based trust, positive values, Good will, biases elimination, ethics-washing, ethical guidelines, marketing communication, Human-related factors (expertise, culture, personal traits), AI-related factors (accuracy |

## 3.1. Trust in Human-Machine interaction: typology and parameters

### 1.1.1    Theories and definitions of trust in AI

Trust is a central component of interaction between people and artificial intelligence (AI) as well as machine and AI since 'incorrect' levels of trust may cause misuse, abuse, or disuse of the technology [18]. To understand trust's implications and influential factors, we first need to have a formalized definition of trust that is expandable to AI. Generally, trust is defined as a directional transaction between two parties. In this definition, A trusts B if it believes that B will act in its best interest and accepts vulnerability to B's actions [32]. Trust is then necessary to predict events by anticipating the impact of actions and behaviors and facilitating collaboration between the parties [33]. Hence, trust is tightly coupled with vulnerability, anticipation, risk, and uncertainty. The trustor needs to anticipate the trustee's behavior to know if it matches its desires. Still, at the same time, the trustor knows that there is a level of uncertainty associated with this anticipation. Therefore, there is a risk of disadvantageous or otherwise undesirable events [18].

The critical question is how to adapt the general definition of trust to the notion of AI. AI can be broadly defined as a computer program that can make intelligent decisions [34]. In the context of AI, the meaning of anticipation in trust changes since the goal of the trustor is not necessarily to anticipate AI's behavior; instead, the trustor needs to anticipate if the model is correct and confident in its decision. This definition can be further expanded based on the theory of contractual trust, which states that the trustor should anticipate or believe that the trustee will stick to a specific contract, which could be any functionality that is deemed useful [35], [36]. In that sense, the former definition would be a specific case of trust in AI, which is the trust in the model's "correctness." In some instances, humans may trust in other functionalities of an AI model rather than its correctness. For example, a classifier trained for medical samples may reveal strong correlations between attributes for one of the classes, demonstrating causation between the attributes, even if the model is not helpful for the original classification task [37].

From a different perspective, while interpersonal trust is associated with benevolence, integrity, and ability, the trust in AI is less relevant to honesty and benevolence since AI systems lack intentionality [38]. Trust in AI heavily depends on the user's perception of an AI system's ability, which depends on the quality of the input data, the mathematical problem representation, and the algorithms used in the decision-making. In general, AI systems could be generative, and they could learn, evolve and permanently change their functional capacities with operational and contextual information [39]. As a result, AI-based systems' actions and decisions could become more indeterminate across time, making them more challenging to predict. Consequently, establishing trust between humans and AI-based systems is generally more complex and challenging to understand than interpersonal trust [40].

A vital phenomenon here is the fact that trust and trustworthiness are entirely disentangled: pursuing one does not entail following the other, and trust can exist in a model that is not trustworthy, or a trustworthy model does not necessarily gain trust [41]. For example, in the healthcare domain, a highly complex classifier trained to identify the risk of cardiovascular diseases from a combination of genetics, lifestyle, and metabolic factors may show a high accuracy, meaning that it is trustworthy regarding the correctness; however, this model may not be trusted by the healthcare providers since the logic behind its decision is not clear. A different example is trust in an untrustworthy AI. For instance, there is a correlative but not a causal relationship between high-quality visual interface (GUI) and trustworthy AI models. If the cause of the user's trust is the model GUI, then the model's ability to make correct predictions will not affect this trust [18]. Even an untrustworthy AI model with poor performance could be trusted merely because of the good GUI. Ghassemi et al. showed a case where the interface can increase doctors' confidence in a tool, despite not significantly increasing the AI's accuracy [42].

It is also shown that, a model is more trustworthy when the observable decision process of the model matches user priors on what this process should be. This is equivalent to, for example, a doctor that is considered more trustworthy because they are citing various respectable studies to justify their claims. The relationship between actual trustworthiness, which is a characteristic of the trustee, and perceived trustworthiness, which is a characteristic of the trustor, and its influence on the trust, has been modeled in [43]. The central assumption is that the actual trustworthiness cannot be accessed directly and is therefore inferred via cues to form a user's perceived trustworthiness. Cues are observable pieces of information such as the aesthetic of a user interface, information about the inputs a system uses, single outputs that a system produces, the displayed predictive power of a classifier, information about uncertainty accompanying a classification output, a stated or communicated rationale for the system's recommendation, or the logo of a company. These cues could significantly affect the perceived trustworthiness and the trust consequently. Therefore, engineering and misleading cues could lead to over-trust.

Accurate assessment of actual trustworthiness that leads to realistic perceived trustworthiness is affected by four factors, including relevance, availability, detection, and utilization, where relevance and availability are associated with the trustee (e.g., an AI system), and detection and utilization are associated with the trustor (e.g., the user) [43]. A relevant cue can be any information regarding a system's predictive accuracy in a task. Relevant cues provide information related to a system's performance, which is considered a facet of trustworthiness [44]. Availability means that only the cues that are accessible to the trustor can be leveraged to assess trustworthiness. Detection refers to the fact that the trustor must detect the relevant and available cues, and the utilization means that the trustors must be able to correctly interpret the relevant, available, and detected cues towards the estimation of trustworthiness [43].

For an AI system to be perceived as trustworthy, five principles need to be fulfilled, including beneficence, non-maleficence, autonomy, justice, and explicability [45]. Beneficence refers to the development, deployment, and use of AI that is beneficial to humanity and the planet and respects fundamental human rights. On the other hand, non-maleficence means development, deployment, and use of AI such that it avoids bringing harm to people. The autonomy principle is not directly related to trusting beliefs, but it helps mitigate integrity and reliability risks by balancing human- and machine-led decision-making. Justice is also a broad term that covers the utilization of AI to amend past inequities like discrimination and biases and the creation of shareable and subsequent distribution of benefits. Finally, explicability requires the creation of explainable and interpretable AI models while maintaining high levels of performance and accuracy from a practical perspective and creating accountable AI from an ethical perspective [19]. Different guidelines proposed for building ethical and trustworthy AI have addressed different combinations of these principles.

### 1.1.2    Trust in types of human-Machine interaction

Recent technological breakthroughs in artificial intelligence and machine learning have generated a surge of interest in the usage of AI technology in daily life. Many applications such as healthcare, autonomous vehicles, financing, and marketing benefit from the development of AI technology. Nowadays, AI is tied to many daily tasks, and different types of interaction between humans and machines occur every day. Figure 2 shows different types of interactions, including human-machine, machine-human, and machine-machine interactions. Herein, we refer to machines as the host of AI technology. Therefore, machines have the capability of reasoning and decision-making based on the data.

**Figure 2: Types of humans, object, and AI trust interaction**

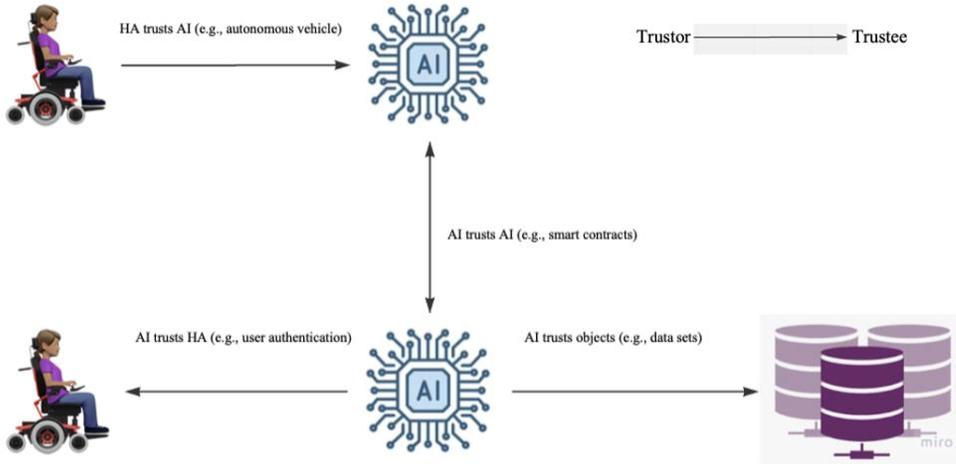

Human-machine interaction is the most common type of interaction with AI, in which the trustor is the human user and the trustee is the AI system. For example, in healthcare, AI could process medical images to diagnose cancer, and the trustor would be physicians who adjust or base their decision on the outcome of the AI model [38]. In another example, Robo-advisors make investment advice [31]. In autonomous vehicles, human drivers interact with the AI-based driver in safety-critical situations [47]. Humans also often use AI-powered personal assistants and chatbots [48], [49]. Examples of human-AI interactions are endless, and trust is essential to facilitate this interaction. Machine-human interaction is a more special case that has not been widely addressed in the literature. The AI systems need to acquire information from human users to update their algorithms. In this scenario, the AI system needs to ensure acquiring data or annotations from trustworthy resources. Adversarial attacks or unreliable sources of information could lead to poor performance of the AI systems. In addition, in many privacy-critical applications such as healthcare, the AI system needs to identify and authorize trusted human users before sharing the data. In this case, user authorization would be essential. Finally, machine-machine interactions are becoming more popular in the light of technology development. For example, sensor networks and IoT (internet of things) strongly rely on machine-machine interactions. In addition, the domains of electronic financing, smart contracts, cryptocurrency, and smart vehicles require an extensive amount of interaction between different AI systems in which trust is paramount since adversarial attacks in these cases are highly possible [50].

Due to the prevalence of human-AI interactions and the fact that it was the first type of interaction since AI has emerged, this is the most widely studied topic in the literature. In particular, factors such as access to knowledge, transparency, explainability, certification, as well as self-imposed standards and guidelines are important to build trust in human-machine interactions. Table 3 summarizes some of the most important factors of trust and their belonging category based on benevolence, integrity, and ability [51]. All these factors matter in promoting trust in human-machine interactions as the human is the trustor. However, in other types of interactions, mostly integrity factors and data governance matter, while transparency and explainability are less important than technical correctness and integrity parameters.

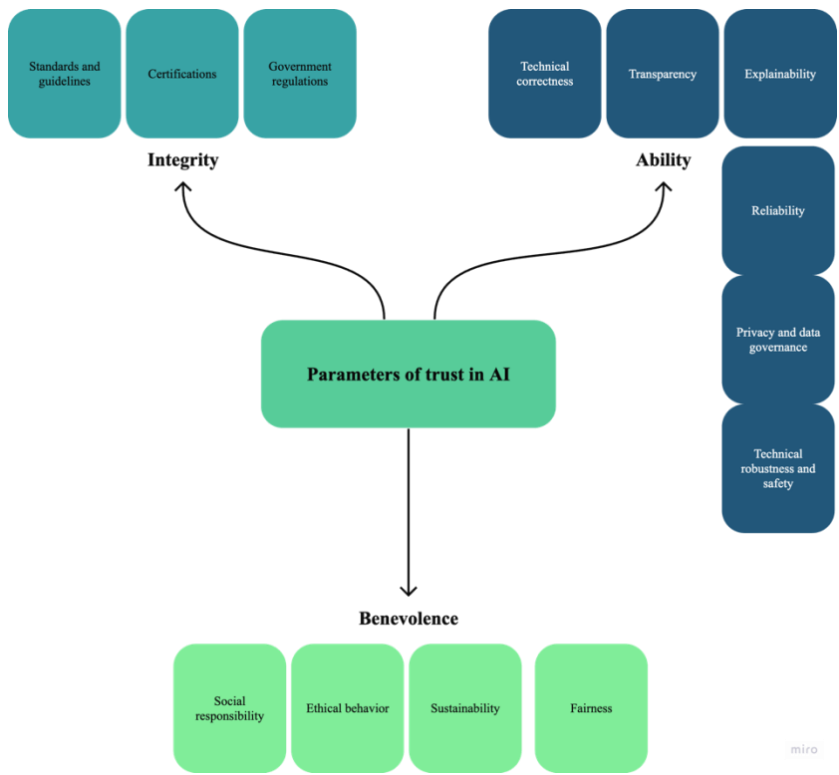

**Table 3: Parameters of trust in AI**

The safety and efficiency of human-machine collaboration depend on the perceived trust of the human trustor. Over-trusting the AI system may cause serious safety issues. A method of adaptive trust calibration was developed by detecting the inappropriate calibration status via monitoring the user's reliance behavior and cognitive cues to prompt the user to reinitiate trust calibration. This becomes significant in applications such as military coalition operations, where data is limited and often of low quality. These problems can be mitigated by taking steps that allow rapid trust calibration so that decision-makers understand the AI system's limitations and likely failures and can calibrate their trust in its outputs appropriately. An AI service can achieve this by being both interpretable and uncertainty-aware [52].

Although considering all these factors could increase the trustworthiness of AI systems, in the case of human-machine interaction, personal traits, especially emotions, play an important role in developing trust as the trustor is a human. It was shown that the humanness of AI applications is an important basis for trusting bonds in human-machine interactions [53]. People may express conflicting concerns about lack of empathy in an AI decision where some may see it as a positive aspect that increases trust in the process by keeping human emotions in check, while others think of AI's lack of empathy and morality disqualify it for making higher-stakes decisions [54]. It was argued that personality often overrides any external influence on trust [55]. Researchers found that different people might have different perceived trust throughout the process of engaging with AI systems [56]. Based on these findings, open individuals trust the AI decision more than non-open individuals. Moreover, trust in AI changes over time when the trustor is a human. In many human-machine interactions, there is a need for collaboration between humans and AI agents. For example, in automated driving, the automated driving agent may release the control for the human driver to take over in certain critical situations. In this case, there are important questions that affect this interaction. First, how should functions between humans and machines be allocated? Answering this question requires technical and contextual knowledge. For example, when the automated driving agent faces a critical situation where it cannot handle or is uncertain about its decision, it should initiate the transfer of control [57]. In another example, when an AI system is not certain about its decision due to noisy data or lack of training, it can interact with the human user to verify its decision or obtain more training [58]. The second question is who is doing the allocation? The AI or the human user? Third, who can authorize an allocation?

There are several risks associated with human-AI interactions, mostly related to the AI agent's performance, fairness, and transparency. In the absence of transparency and explainability of the AI, the human does not have enough information to form a judgment regarding the chosen decision. Moreover, undesirably biased recommendations could make humans accountable for unethical or legally uncompliant decisions [59]. Bias can exist in many shapes and forms [60] – data-related biases such as measurement bias, historical bias, population bias, longitudinal data fallacy, and social bias, algorithm-related biases such as linking bias, omitting important and influential variables from model, and algorithmic biases, and results interpretation-related biases, such as aggregation bias, user interaction bias, and evaluation bias.

AI-AI interaction is an emerging paradigm. For example, smart and connected vehicles have gradually stepped into our daily lives, and they generally rely on vehicular networks to generate and exchange traffic-related messages. Malicious accidents could result from the untrusted content of vehicle navigation and autonomous system. Blockchain and artificial intelligence (AI) empowered trust management systems were developed for trust evaluation, where it was leveraged to filter the information gained from other smart vehicles (Pan et al., 2020; Zhang et al., 2021). In another study, an Intelligent Trust Collaboration Network System (ITCN) was developed to collect data through collaboration with mobile vehicles and Unmanned Aerial Vehicle (UAV), in which there is a score determining trust associated with each vehicle [63]. Distributed systems of software agents are another example of AI-AI interaction where the AI agents cooperate in helping their users to find services provided by different agents. Examples are prevalent in blockchain, cryptocurrency, and smart contracts [64], [65], [66], [67], [68]. In this scenario, the agents need to ensure that the service providers they select are trustworthy. Because the agents are autonomous and there is no central trusted authority, the agents help each other to determine the trustworthiness of the service providers they are interested in. A trust network is a multiagent system where each agent potentially rates the trustworthiness of another agent [69].

AA-Object interaction is the final paradigm; however, it has barely been discussed in the literature so far. For instance, a self-driving vehicle's AI system needs to trust stop signs; or an AI system, which is used in disaster management requires recognition of un/trustworthy social network data sets during disasters. In addition to physical objects, a trustee, in this paradigm, would also include mental objects, such as theories, thoughts, and algorithms that an evaluative AI system must handle. There are also complex objects at work such as institutions and systems, as trustees.[70]

### 1.1.3 Impact of trust/distrust on AI technology acceptance in different domains

AI is one of the most-discussed technology trends in research and practice today and is estimated to deliver an additional global economic output of around 13 trillion dollars by the year 2030 [71]. Various domains benefit from the AI as it helps decision-makers and provides vital services for end-users. AI has significantly affected domains such as healthcare [72], [73], [74], [75], finance [76], [77], [78], personal assistance [79], [80], [81], autonomous vehicles [24], etc. The importance of trust in AI extends to other areas as well. Electronic markets, for example, are increasingly augmented with AI-based systems such as customer service chatbots [82]. Likewise, several cloud providers recently began offering

"AI as a Service," referring to web services for organizations and individuals interested in training, building, and deploying AI-based systems [83].

Trust is paramount for the well-functioning of healthcare systems and, consequently, for the acceptance of AI by physicians and within healthcare more broadly [84]. Transparency and explainability are the most important factors of trust in healthcare systems [41], [85]. Other concerns that can decrease physicians' trust in AI include, among others, the low number of randomized clinical trials to test the performance of AI systems, the lack of transparency of information flows within AI applications, the risk of inequity and discrimination introduced by algorithmic biases, and insufficient regulatory clarity [86], [87]. In addition, limited public literacy about AI negatively affects trust in healthcare [41].

In a study investigating the trust of the end-users in detecting heart diseases from ECG signals acquired by smartwatches and detecting cancer from skin photos, it was found that several participants appreciated the more holistic perspective of a human doctor compared to the limited focus of an AI-powered app. Moreover, they mentioned the social impact of visiting and talking to a human doctor, which is missing in AI-driven apps. Many participants raised concerns regarding the overall technical feasibility of these AI-driven diagnostics, and they did not want to be notified about life-threatening health complications through an app. Finally, although some participants found the AI-driven diagnostics helpful to reduce the wait time, they mentioned that they would first test the performance of such an AI-based app themselves [88]. These findings show the distrust in AI-driven diagnostics systems, leading to lower acceptance of technology. In order to replace or supplement human diagnosis from physicians and health care professionals, it may not be enough for the AI diagnosis system to be just accurate as an accurate diagnosis without justification or explanation might be ignored. The format and the timing of explanation play important roles in regulating trust in healthcare systems [89]. Algorithmic analysis of pathology images is one of the most promising and advanced applications of AI in healthcare [90]. However, few AI systems are currently being used in the field, and it is uncertain to what extent pathologists will adopt AI and rely on its recommendations.

Another example of the role of trust as a driving factor for technology adoption is found in the financial sector. While AI already enjoys a high level of trust in some areas (entertainment, navigation), only half of people trust algorithmic investment advice [46], [91]. Many human investors would rather trust a human prediction than an algorithmic prediction [92], a phenomenon known as algorithm aversion because humans are more tolerant if a human is mistaken than if it is an algorithm. When relying on AI algorithms to manage investment, humans' loss tolerance is highest when humanized algorithms manage portfolios – e.g., giving the algorithm a human name [46]. Several banks have leveraged chatbots for interaction with customers in the financial sector. One of the advantages of AI systems to gain customers' trust is the inherent absence of self-interest. Nevertheless, humans are still preferred to advise customers concerning complex financial products such as equity derivatives. Humans are also preferred when customers wish to complain or discuss a complicated matter or situation. A common criticism of chatbots and robots is that they cannot empathize [89].

AI-based personal assistants, chatbots, and coaches are other domains in which trust in AI directly impacts technology adoption. AI-based voice-assistant systems (VAS) are used for various purposes in daily lives. It was found that interaction quality (e.g., information and system quality) and trust are critical factors influencing the adoption of AI-based VASs [93]. Although the current AI systems do not have an internal drive to misbehave, lack of transparency in these systems may seem to indicate deception to some users [94]. For example, Microsoft's AI bot, "Tay," was meant to learn to chat by communicating with inter-net users. Instead, due to toxic influences, it began spouting racist and genocidal ideologies, which resulted in distrust and users' outrage that forced Microsoft to suspend the system [95].

Despite the significant potential of AI in the manufacturing industry, its application still faces the challenge of insufficient trust. Research on how users trust AI in an organization such as a manufacturing company is rare. In the organizational context, the decision of trust in AI is not completely personal; instead, users must consider the institutional influences of the company, the leader, or peers before they make the final trust decision. Research showed that factors such as gender, age, education, and position had no significant effect on organizational trust. Instead, the support from the top manager acts as the endorsement to ensure that the AI is qualified and that the AI-related project will be successful, which enhances trust in the organization [96].

Robotics is another important field empowered by AI, which requires trust between humans and machines. Influential factors of trustworthiness in the context of social robots were investigated in [97]. Robot relevant issues (e.g., the characteristics and performance of the robot), human-relevant issues (the specific need, propensity to trust, personality, comfort, self-confidence, attitude, memory, attention, expertise, competency, workload, prior experience, and situation awareness), and scenario relevant issues (task application, task complexity, multi-tasking requirement, physical environment, in-group membership, culture, communication, team collaboration, etc.) were found significant factors, among which robot-relevant issues are the most significant factors influencing people's trustworthiness evaluation towards human-robot interaction. In conclusion, it was shown that cognitive trust and emotional trust are positively related to the intention to adopt an AI-based recommendation system as a decision aid, where cognitive trust has a stronger effect. Moreover, emotional and cognitive trusts were found correlated [98].

## 3.2. Trustworthy AI and its metrics: Technical, and non-technical metrics

The increasing use of artificial intelligence (AI) in various industries, including healthcare, has raised concerns about its trustworthiness. Trustworthy AI is critical for ensuring that AI systems are reliable, safe, and ethical. In this context, several technical and non-technical metrics have been proposed to evaluate the trustworthiness of AI systems in healthcare. Focusing on many clinical research in AI and robotic abilities for diagnosis of diseases or rehabilitation assistance revealed the significance of discussion about trust definition in the clinical decisions or suggestions and factors that can improve the clinicians and technical trust to the AI [99], [100] [101]. In addition, it is needed to focus on some non-technical (e.g., ethical or legal) foundation for autonomous AI requirements (such as maximizing traceability of patient and ongoing monitoring of real-world performance), which can be found in [102] and references therein. On the other hand, there are some articles that clarify why we can't use AI in medicine as a trusty system [103], and/or use it with some limitations because AI reliability is insufficient [104] that can enable better interpretation [105]. In categorization which was carried out by Kush R. Varshney [106], accuracy and safety are preliminary metrics for AI, reliability includes fairness and robustness, and also, transparency (which includes interpretability and explainability) and value alignment are essential for human-AI interaction. He also itemized the principles in ethics guidelines as 1) privacy, 2) fairness and justice, 3) safety and reliability, 4) transparency, and 5) social responsibility and beneficence.

There are some factors in the other industrial domains to achieve trust such as using standard definitions, a system for complaints declaration, and independent rating services [107]. Since trust in new technologies and AI is one of the human concerns, many companies and agencies established formal validation and verification of autonomous robot's software [108], surveyed trust metrics and modeling [109], [110], [111], and provided Robotics and AI Roadmap [112] ;[113];[114]; [115]; [116] to carry out about algorithmic ethics and human-AI interaction in the field of autonomous robotic systems. Furthermore, [117] provided more information about ethics guidelines and approaches employed in European survey. A comprehensive research about trust and relationships in AI studied robot imitation, understanding and AI-human interaction concerning concepts in biology, neuroscience, social psychology and sociology using outcome matrices as a tool of robot's interactions [118].

In what follows we elaborate on 7 non-technical and 3 technical metrics for trustworthy AI, as well as some measurement models and frameworks for trust/trustworthiness in AI.

### 3.2.1. Trust & explainability / transparency / interpretability

In solving complex problems, most AI methods are based on the direct use of complex opaque concepts such as deep neural networks [119]. There are some metrics to measure the performance of AI models, such as testing the model on a test set or cross-validation score. However, in most complex AI-based solutions, even the developer has limited access to the mechanism in which the model processes the input data. This opaque nature of complex AI algorithms in turning the input into output is referred to as "black-box" AI [120], [121], [122]. The trustworthiness of these algorithms has been questioned by many ethical, technical, and engineering communities [120], [122]. The pervasive use of deep neural networks in which the number of input features sometimes exceeds thousands of nodes has exacerbated these concerns [123]. Accordingly, AI scientists in recent years have focused on a branch of AI called Explainable AI (XAI), which aims to add explanation, transparency, and interpretation to AI-based decisions by shedding light on the opaque nature of AI methods [124]. Studies have shown that XAI can increase the trust of the end-user in AI-based decisions [125].

Transparency is one of the fundamental ethical principles in creating trust in users toward AI decisions [3]. Although transparency and explainability have been usually categorized under the same ethical principle [126], it is essential to distinguish between these two different topics before extensive interchangeable misuse of them. Explanations seek broader goals, and transparency (explaining how clear the system reached the answer) is one of them [127], [128]. Research studies have shown that transparency averts overtrusting AI [129]. However, other types of explanations, such as justification, might lead to users' overtrust by representing manipulative information [130]. Also, researchers have warned that too much focus on transparency, especially at the early stages of an AI product, can damage innovations [131]. In addition, it is worth mentioning that different stakeholders look for different facts in an AI model. Thus, the level of transparency reported to different stakeholders might be different [132], [133]. In [133], the authors divide the stakeholders into five big categories, including developer, regulator, deployer, user, and society in general, and talk about how much detail of transparency they are looking for. Sometimes, even the required level of transparency within one category of stakeholders might be different, for instance, depending on their social geography [134] or their personality [135]. As a result, defining context-based transparency criteria is difficult to achieve [131].

Building trust is dynamic [136], ranging from initial trust to ongoing trust [137]. Needless to say, explanations are a hand-in-hand partner in this dynamic process [127]. However, the pervasive prevalence of using opaque deep neural networks in AI in recent years has challenged the explainability of the models and, thus, the perceived trustworthiness of the users. These black-box networks are complex and opaque in terms of operation [120], and even sometimes, the developer has limited access to how they operate. Their complexity is also the underlying reason for their outstanding performance in outperforming conventional solutions and other AI models [138]. Thus, there is a tradeoff between the desired accuracy and the level of transparency and explainability, meaning that models with the clearest explanations, such as decision trees, may not have a good performance, while those that are the most accurate, such as deep learning based models, are the least explainable [139], [140]. XAI models have been given momentum recently to mitigate this tradeoff and to open the black box by providing transparency and explainability to AI-based models [141] in different areas, including medical domains [142], [143], robotics [144], autonomous transportation [145], and stocks [146], [147], [148]. Recent data regulation set by European Union (EU), known as General Data Protection Regulation (GDPR), has attracted more attention to this field. GDPR recognizes the right of EU citizens not to accept decisions made solely based on automated processing [149], which incentivize XAI.

Interpretability is another crucial aspect in increasing users' trust in AI-based decisions [150]. It should be noted that interpretability and explainability share some common goals, yet they are two different things and should not be used interchangeably [151]. Interpretability aims at making the decision's rationale understandable for the stakeholders, i.e., the relationship between the cause (input) and effect (output) is human-interpretable [152]. But explainability is a deeper concept that is concerned not only with the system's inference but also with the model's inner machinery, i.e., how the model works and the way the model is trained [153], [154]. Thus, interpretability can be categorized as a subset of explainability [155]. Kamath *et al.* [156] further categorize the interpretability of XAI methods into three stages: a) pre-model interpretability, b) intrinsic interpretability, and c) post-hoc interpretability.

Pre-model interpretability emphasizes the importance of understanding the data set through exploratory data analysis, data visualization, and feature engineering before model selection [157], [158], [159] and the fact that there is no "go-to" model that fits all data sets. Intrinsic interpretability refers to techniques that are intrinsically interpretable due to their structure [160], [161], [162]. It ranges from basic models such as decision trees to advanced ones such as explainable boosting machines. Post-hoc interpretable methods refer to methods that utilize the power of complex black-box models in accurate predictions and try to add global or local interpretability to their decisions [163], [164], [165], [166]. Similar to transparency, some researchers proposed that the level of interpretability of a system should depend on the category of the entity working with the system, e.g., operators, executors, examiners, and the system should reveal a set of suitable measures of interpretability based on their relation to the system [167]. Besides, a consensual definition for interpretability in AI and how to quantify it has not yet been reached [152], [168], [169]. Despite these challenges, it is hoped that interpretability will soon reach a state of readiness[170].

### 3.2.2.        Trust & empathy in AI

Empathy is often considered a crucial factor in building trust in all cases, particularly in relation between human users and AI systems (e.g., see Gamer et al, 2010). It is defined as a subjective process in which one comes to have a deep understanding of other people's feelings [171], particularly in non-judgmental way (Wiesman 1996), or an "ability to simulate how others subjectively experience a situation and how they regulate elicited emotions" [172]. In fact, empathy involves two main components: the cognitive ability to make an accurate inference of what others think and feel (empathic accuracy) and the affective or emotional ability to make a supportive, benevolent, and compassionate response to their thoughts and feelings [173]; [174]. A distinction has been drawn between cognitive and behavioral aspects of both trust and empathy, where the latter involves the relevant agent's behaviors, such as a robot's ability to safely lift a patient, and the former involves the agent's cognitive abilities such as the ability to provide accurate information, make proper inferences, or exhibit an understanding of the other's feelings and thoughts. In these two aspects, trust is essential for an efficient communication, and empathy is key to social bonding, which facilitates social interactions, and hence, helps build trust between people, and between persons and AI systems, since social agents are more trusted if they show an understanding of people's mental states [172]. Moreover, empathy has links to expectation: if the AI system can act in accordance with its user's expectations, the user will probably form a trust in the system [172]. Empathic accuracy, as an essential element of empathy, has an impact on the agent's credibility and trust [175]. Gebhard and colleagues [172] have identified the following requirements for "empathic cultural-aware agents"; that is, those that take social values and norms into consideration: explainability on both behavioral and motivational levels; observation and detection of social signals such as smiles, facial expressions, and postures; interpretation of utterances and simulation; empathic action and interaction (showing respect for the values and norms of others); adaptability to individuals (showing respect for individual aspects such as their levels of hearing or their dialects). Empathy also has an indirect role in trust through accountability: an agent can be trusted if, among other things, it is accountable, and accountability requires consideration of the stakeholders' viewpoints and needs; that is, empathy [176]. Another significant link between trust and empathy is established by findings about user similarity and trust. For example, users tend to trust online recommendations based on preferences by other users with similar tastes, which are extracted from similar ratings or online purchases and the like [177]. Similarly, it is found that users tend to trust agents with values similar to their own [178]. The necessity of transformations in the notions of empathy and trust in patient-doctor relations in the age of AI-based treatments or patient online communities has been highlighted ([179] [180], [181]). The link between empathy and trust in the case of online or AI-based services is discussed in Bock et al, [182] and Yoon and Lee [183].

### 3.2.3. Trust and privacy

There are obvious tradeoffs between trust and privacy. The idea is articulated in terms of negative associations between online privacy concerns and trust ([184]; [185];[186];[187]: the higher privacy concerns the lower trust [188]. Privacy is defined as self-determination of when, how, and how much one's personal information or personally identifiable information is communicated to others; that is, privacy is one's control over identity privacy (information leading to identification of a specific person), location privacy (information from which one's location can be identified), communication privacy (confidentiality of one's information), access privacy (control of access privileges), and data processing privacy ("information about the information flow in processing and dissemination of data")[189]. Control over personal information is deemed important in many definitions of privacy [190]. In addition to privacy concerns, people's assumptions about their self-efficacy in protecting their data play a role in their trust in AI [191]. To provide better services and to keep and attract customers, AI companies need a plethora of information, which depends on detailed information from their customers, but this raises privacy concerns on the part of the customers [190]. As a result, they lose their trust, and as a consequence, they become reluctant to share detailed or accurate personal information, which in turn lessens the value that the companies were supposed to gain from personal information [186]. Moreover, this tends to result in avoidance of online shopping ([190]; [192]). An increase in trust leads to increased and more accurate information sharing and decreased perceived risk [188]; [193]). Van Dyke and colleagues propose "customer privacy empowerment" to increase trust and encourage information sharing. The idea is to give customers greater control of how, when, and how much personal information is used ([190] [194]). The European Community has published the General Data Protection Regulation in 2016, according to which service providers are required to answer user questions about the location of the data, whether their information might be read by others, whether their information is traced, or whether they can revoke permission to use their personal information [189]. This could be done through transparency on the part of the producer or service provider by giving information on the lineage and providence of the product [195], by giving explanations of the product or service [196], or by social presence and social attributes of the AI system, as in voice assistants [197]. This is an external legal guarantee for protection of privacy, but "privacy-by-design" embeds privacy requirements in the system's design; e.g., by data minimization, controllability, transparency, easy-to-use privacy function, data confidentiality, technical quality of data, and limited use of data [189]. A significant point here is that since varying degrees of trust are needed in different contexts, different levels of privacy (high, medium, and low) might be required[189]. Technical solutions have been proposed for the tension between privacy and trust [193], such as secure two-party computation techniques based on homomorphic encryption [198], certain blockchain implementations [199], a model that only warrants cooperative AI systems to receive high-fidelity information [200], and anonymous authentication and attack tracking [201]. Some studies show that people are more likely to share information when the technology in question offers a much needed function or is pleasurable ([202]; [197] while they are reluctant to do so when it comes to scientific surveys and interviews. Richards and Woodrow [203] suggest that extant privacy laws have "a pessimism problem" in that they excessively focus on harms from privacy infringements and make too much of people's ability to opt out of possibly harmful data practices. In contrast, they propose that privacy should be seen as what enables trust in major information relationships, in which way value is created for all parties to an information exchange by establishing a sustainable data relationship.

### 3.2.4.    Trust and fairness in AI

Algorithmic discrimination and bias tend to hinder the trust in AI, while perceived fairness or justice enhances trust [204], [205]; that is, to trust a system users should be assured that it can act justly or in an unbiased manner toward all groups [206]. Fairness in AI is equal treatment or equitability of an AI decision about various groups of users [205]. Algorithmic unfairness, in many cases, caused by failure to develop AI systems based on a fair training of data or a fair design of the relevant machine learning model [205]. For instance, in AI-based medical diagnostic systems discriminations and biases might arise when little or no information from black-skinned and other minorities is fed into the system during its development [207]. When it comes to perceived fairness and trust, it was found that people see human decisions fairer and thus more trustworthy than algorithmic decisions in the case of human (as opposed to mechanical) tasks [208];[209]. This is not true, however, of minority and marginalized groups: they tend to perceive algorithmic decisions as trustworthy as human decisions [210]. Moreover, it was found that perceived fairness is positively related to induced fairness, which is to say that a high degree of induced fairness culminates in a high degree of perceived fairness by people, and the latter is in turn positively related to user trust [205]. User biases might also affect their perceptions of trust in an explainable AI system: differences in user trust have been found between malignant and benign diagnoses of an AI system [211]. It is also found that perceptions of harm and injustice as well as reported wrongdoing are positively related to uncanniness, which in turn negatively influences trust in an AI agent [204]. Implementation of rules and regulations is deemed necessary for achieving fair, trustworthy AI systems [212].

### 3.2.5.　　Trust and accountability in AI

Contemporary research on accountability in AI and machine learning is mainly focused on defining the rights of human stakeholders, obligations of developers, and ways to enforce them. This approach, known as "offloading," has led to the central concept of the "right to explanation," which demands that AI systems provide justifications for their actions. This focus on accountability as answerability has led to the development of a regulatory framework and a system design approach that ensures that the AI system can provide the right kind of answers. Thus, the primary aim of accountability in AI and machine learning research is to define the rights and obligations of stakeholders and to build AI systems capable of providing satisfactory explanations for their actions.

Previous research has identified the need for a robust legal framework for establishing and maintaining trust in artificial intelligence [213], [214], [215]. While one aspect of public trust in AI is the reliability of models and the individual recommendations of those models, willingness to trust (and the underlying trustworthiness that willingness tracks) is situated in a context of public trust in institutions [213]. This suggests a two-pronged approach in which researchers work to improve trust in individual models and recommendations and also work to develop a system of minimum standards, verification, and accountability. With regards to the first prong (that of trust in models and recommendations), one component is developing standards of explanation [216]. Transparent explanations and accountability are a prerequisite for trust in individual decision recommendations. [217][217][217]

The primary focus of contemporary research in accountability in AI research is on offloading questions. That is, the question is one of clarifying exactly what rights human stakeholders have, what obligations AI developers have, and how governments, developers, and watchdogs can enforce this scheme of rights and obligations [217], [218], [219]. Within this project, the so-called 'right to explanation' has been central [217], [218], [220], [221], [222], [223], [224], [225], [226], [227], [228]. In one sense, this focus makes sense for a discussion of accountability. It has been argued that accountability is either, at its core or in part, a matter of answerability [229], [230]. If we take the demands of answerability literally in this way, then an accountable system of artificial intelligence will justify its actions [229]. We can then proceed from the offloading project of accounting for a right to good explanation (whatever we take that to mean) and a regulatory framework guaranteeing it to the agent-building project of designing a system capable of providing the right kind of answers.

### 3.2.6.    Trust and technical metrics (safety, accuracy, robustness)

In [231], a precise discussion is presented regarding the nature of trust in AI, as well as the prerequisites and goals of the cognitive mechanism of trust. Their model, based on interpersonal trust, considers both the vulnerability of the user and their ability to accurately assess the impact of AI decisions. Several technical aspects of trust are proposed to be addressed in AI systems, including reliability, safety (encompassing fairness and explainability), security, and lineage [232]; [107]. Other studies have shown that providing human-meaningful explanations regarding the system's accuracy can influence user understanding and subsequently enhance trust in AI performance [233], [234]. Additionally, these findings suggest that accuracy is a more influential factor than explainability when it comes to improving user trust [235].

When it comes to making high-stakes decisions, particularly in fields such as law, medicine, and the military, trust and reliance on AI systems become more challenging. In order to address this, [236] explains the concept of trust calibration, which involves making AI systems interpretable and uncertainty-aware. By incorporating interpretability and awareness of uncertainty, trust in AI systems can be better calibrated. Another model, as described in [237], compares the decisions made by AI systems and humans in their respective tasks to determine when to trust or distrust the AI. This model helps establish guidelines for understanding the appropriate level of trust to place in AI systems. Additionally, [238] presents an adaptive trust calibration approach for human-AI interaction to analyze instances of over-trust in AI.

In [239], a novel sparse decision-making model is proposed that integrates trust and information rating. This model takes into account both trust and the quality of information when making decisions. Several articles propose various mechanisms to increase trust, such as supplier's declaration of conformity (SDoC) for AI services [232] or the use of FactSheets [240], which are filled out by both AI service providers and users. These mechanisms aim to enhance transparency and accountability, thereby fostering trust in AI systems.In the context of trusting the evolution of 5G internet services, a conceptual zero-touch security and trust architecture has been proposed [241]. This architecture aims to ensure secure and trusted communication in the 5G network. Additionally, it has been suggested that combining diversity (utilizing network nodes with different characteristics) and trust (immunity from failures and attacks) can enhance the structural robustness of sparse networks [242].

To address trust and knowledge sharing in graph models, a blockchain-based approach has been introduced [243]. This method facilitates the sharing of trusted knowledge, isolates malicious nodes, and prevents knowledge pollution, thereby promoting reliable information exchange. The utilization of AI has demonstrated increased efficiency in various tasks [e.g., Rahman et al. [244], [245]]. However, it should be noted that trust in AI is not guaranteed in the realm of cybersecurity. Nonetheless, it is argued that trust can play a role in improving the design, development, and deployment of AI systems [246].

Several articles have developed structured models with mathematical definitions to explore various aspects of trust. These models include parameters such as trust system space, maximal and intuitive attacker models, and robustness properties [247]. Other models focus on advisors hiding or minimizing their true observations [233], a Bayesian-based trust model for human multi-robot teams [248], and factors related to both humans and robots [111] to discuss and formulate trust robustness. In [249], a psychological metric called the weight of advice (WoA) is employed to analyze human-AI interactions when advice is provided by both human and AI sources. The study reveals that participants' behaviors are similar in both cases, but the level of trust varies depending on the topic of the advice. Another study aims to evaluate human trust in AI by examining the relationship between psycho-physiological states, such as biosignals or physiological signals, and trust and cognitive load [250]. The study explores how physiological indicators can provide insights into the level of trust individuals have in AI systems. Overall, these studies contribute to the understanding of trust by utilizing mathematical models, psychological metrics, and physiological signals to examine the robustness and dynamics of trust in various human-AI interactions.

### 3.2.7.     Evaluating and measuring/ trustworthiness certificate in AI

To assess the trust between humans and AI and establish accuracy and safety guidelines for AI-assisted decision-making, numerous psycho-physiological approaches (e.g., [251], [252]) and empirical approaches (e.g., [253], [255]), supported by theoretical methods, have been proposed. These approaches are aided by the use of questionnaires, experimental protocols, qualitative evaluations, and other evaluation techniques. However, several challenges can affect the validity and accuracy of these investigations. Firstly, it may be difficult to encompass all trust factors within questionnaires, experimental protocols, and qualitative evaluations. Additionally, the diverse designs and models of trust, coupled with the dynamic nature of trust influenced by experimental constraints and the methods employed by the AI system, present challenges in developing comprehensive guidelines and protocols [255] [256]. The complexities surrounding the evaluation of trust are further explored in [257], offering insights into the associated problems.

In [253], a trust-theoretical model analyzes consumer trust in mobile payment (m-payment) services, shedding light on user trust in m-payment systems. The verification and validation of autonomous systems are debated in [259], and [260], with a focus on transparency and sharing awareness between designers, testers, and users in the development of transparent AI systems. Cho et al. [260] identify key attributes of trustworthiness (such as reliability, safety, resilience, and agility) in relation to trust. The framework of ontology-based trustworthiness considers vulnerability, errors, and the relationships between these factors to establish a threshold of confidence for AI systems.

In [261], the quality evaluation of computer-based systems is conducted using the aforementioned metrics, incorporating vulnerability and risk assessments. The study aims to identify future research directions and enhance the metrics and methodologies employed. The theoretical aspects of trust in AI within the manufacturing industry are explored in [262], which categorizes trust into three levels: organization, group, and individual. These levels involve factors such as management commitment, authoritarian leadership, and trust in AI promoters.

A psycho-physiological model for assessing user trust in AI is proposed by Ajenaghughrure et al. [263], seeking to determine which user signals provide accurate assessments of trust. In evaluating trust in human-AI interaction, (Schmidt et al., 2020) find that participants prefer physical interaction and embodiment with AI rather than relying solely on voice control. Another study introduces multi-dimensional metrics, including user satisfaction, to assign a trust score to an AI system. This trust score encompasses factors such as job efficiency and effectiveness, understanding, control, and data protection [265].

Hoffman et al. [255]  discuss trust scales in AI and emphasize two key aspects: trust in the output and reliance on machine advice. They provide a comprehensive review of various trust assessment scales and suggest that recommended scales should focus on predictability, reliability, efficiency, and believability of AI systems. [252] employs behavioral and physiological measures, such as individual and team performance scores, team situation awareness, and process measures, to evaluate human-AI interaction and trust in AI. The study reveals that trust levels differ between human-human and human-AI interactions, and interestingly, trust scores for human-human interaction increase in degraded scenarios, in contrast to human-AI interaction.

[266] conducted an influential study that offers comprehensive and well-structured explanations regarding the philosophical evaluation of trust. The research puts forth a rational argument stating that trust in AI is impossible due to its complexity and inexplicability. However, the study highlights the importance of value-based trust, which can be derived intuitively from the information obtained through decision-making algorithms and their implications, such as diagnosis accuracy and safeguarding. The significance of AI certification is also discussed, emphasizing the inclusion of ethical principles and mandatory conformity assessment. This certification process aims to enhance algorithmic auditing, facilitate customization of AI certification, and establish educational

programs addressing AI and its safety concerns. Applying ISO standards to the quality and security management of AI in specific processes is seen as a valuable approach to ensure AI's reliability and safety [267].

### 3.2.8. Trustworthy AI Frameworks

Banavar [268] developed a framework centered around secure and morally sound AI systems that aim to cultivate trust through repeated interactions. Nonetheless, he emphasized the importance of algorithmic accountability, adaptability, precise integration of data, algorithms, and AI systems, as well as safeguarding privacy within this broader understanding of trust. Given that the ethical framework is crucial for human-AI interactions, as discussed in [267], [269] has adopted this framework to examine a model for human-robot interactions. It was emphasized that AI decisions are influenced by human judgment. The relationship between ethics and AI can be summarized into three aspects: ethics by design, ethics in design, and ethics for design [270]. In a study by (Jobin et al., 2019), five ethical principles—transparency, justice and fairness, non-maleficence, responsibility, and privacy—were highlighted to encourage a global convergence in integrating AI system guidelines. Additionally, another related study proposed five foundational principles—beneficence, non-maleficence, autonomy, justice, and explicability—to develop a data-based framework for trustworthy AI [272]. The ethics of AI in global health, as explained in [273], revolve around the metrics of explainability, algorithmic bias, and trust, raising important questions regarding value, fairness, and trust. For a comprehensive review of AI ethics guidelines, the practical implementation of AI and ethics, and advancements in AI ethics, interested readers are referred to [274].

Through an examination of various machine learning approaches in air traffic management, researchers in [275] devised an explainable framework aimed at enhancing trust in AI. Their automated method operates by leveraging existing guidelines and incorporating user feedback to bridge the gap between research transparency and practical explainability. In a separate study, (Shaban-Nejad et al., 2021a) focused on the explainable AI framework in public health and medicine domains, emphasizing the metrics of fairness, accountability, transparency, and ethics. Furthermore, these four factors are deemed crucial in obtaining social license and fostering trust in data [277].

A computing architecture was developed in order to explore the similarities between human and AI decision-making processes using an effective trust model [278]. Within their theoretical framework, the Naïve Bayes method was utilized to classify final decisions, taking into account behavioral patterns derived from human-AI interactions. Furthermore, a theoretical framework was presented to generalize trust antecedents in AI-based conversational agents across various contexts [279]. In industrial settings, a normative framework known as the "invitation of trust" was employed, considering cultural factors and focusing on conversational AI [280]. Autoregressive models were applied to extensive datasets, encompassing dialogues and negotiations, to facilitate an ethical and automated training process for AI chat systems. Trust in technology and its acceptance were the subjects of a theoretical study that aimed to address the risks associated with AI usage, emphasizing the importance of user training [281]. Additionally, the National Institute of Standards and Technology (NIST) is currently engaged in developing a risk management framework specifically for artificial intelligence (AI) [282]. The objective of this framework is to provide organizations with a set of best practices and a shared language for effectively managing risks associated with AI throughout the entire organizational life cycle.

In a conceptual framework presented in [283], the authors initially addressed the issue of trust in AI systems when the processes involved are not adequately understandable and traceable. They then conducted an analysis using two different datasets related to urban logistics planning and heart arrhythmias. The purpose of this analysis was to demonstrate how the identification of patterns can enhance trust in human-AI interactions, emphasizing the importance of examining the results and conducting thorough inspections.

The explainability and trustworthiness of AI pose numerous challenges, making it difficult to make decisions regarding the acceptance of outcomes, as discussed in [284]. To address this, the paper proposes three scenarios: tracking contacts, analyzing big data, and conducting research during public health emergencies. These scenarios aim to establish a consent-based trustworthiness process. Additionally, the framework proposed in [285], [286] is designed to operate based on trustworthy metrics mentioned in the preceding subsection.

In [258], a concise overview of different frameworks addressing the differentiation between human labor and AI labor is provided, along with the importance of understanding how AI systems operate. Furthermore, [287] offers a brief review of the principles outlined by the European Union for trustworthy AI, summarizing the approaches and requirements for establishing trustworthiness in such systems. A redefined multi-level framework for robot autonomy in human-AI interactions is presented in [288], aiming to provide guidelines on how different levels of robot autonomy can impact variables such as acceptance and reliability. [289] argues that instead of trust, the concept of reliance should be used when referring to AI, as AI lacks emotional states and responsibility for its actions.

Within a sociological framework, [231] introduces a trust model that revolves around two key factors: the vulnerabilities of the user and the ability to predict the consequences of AI decisions. The article defines concepts such as contractual trust, warranted trust, and unwarranted trust, offering a formalism for designing trustworthy AI that is based on warranted trust.

The European approach towards establishing global trust involves the development of ethics guidelines within a regulatory framework for AI. These guidelines aim to create an ecosystem of trust that encompasses policy, fundamental rights, and consumer rights. They provide a comprehensive description of seven key factors that are considered crucial in ensuring trustworthy AI: 1-Human agency and oversight, 2-Technical robustness and safety, 3-Privacy and data governance, 4-Transparency, 5-Diversity, non-discrimination and fairness, 6-Societal and environmental well-being, 7-Accountability [290]. To implement these trust measurement approaches, various frameworks have been proposed, which can be visualized in Figure 3. These frameworks serve as guidelines for effectively incorporating and measuring the aforementioned factors of trust in AI systems.

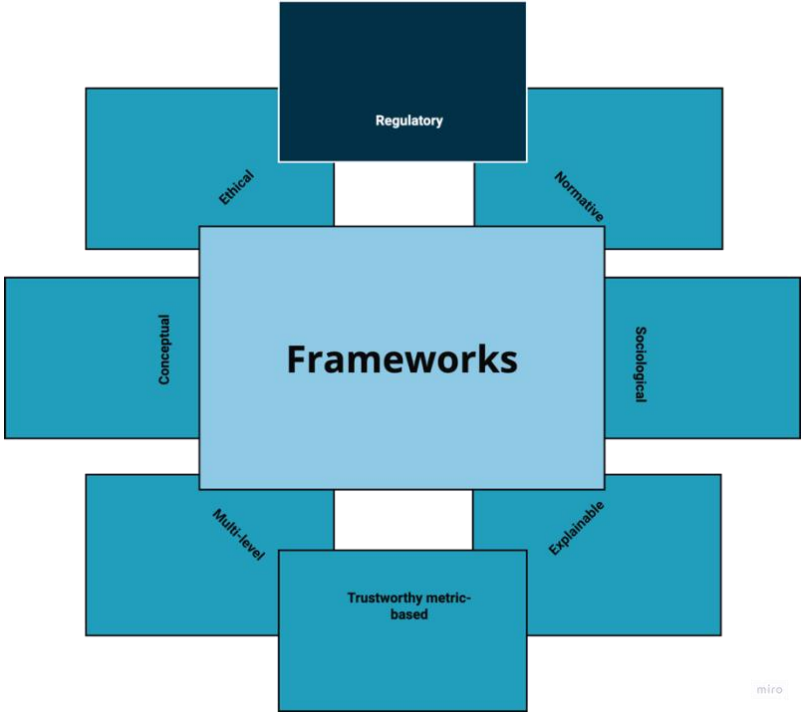

**Figure 3: Different frameworks that can be employed in trustworthy AI.**

### 3.3.    Distrust in AI and Scary AI

Distrust in artificial intelligence tends to be attributed for a variety of reasons. At the macro-level, there are general fears about the power of a machine with artificial general intelligence (AGI) [291], [292], [293]. AGIs are hypothetical artificial intelligence systems possessing broad plastic intelligence (like our own) as opposed to task-specific algorithms. These concerns worry that because an AGI would be able to process information faster than its human counterparts and could have access to the full domain of human knowledge available on the internet, they would be able to outcompete their human creators [292]. Outside of fears of a malevolent artificial intelligence, however, others worry about the so-called 'alignment problem' of how we, as a society, and researchers working on artificial intelligence research could ensure that an AGI's interests and values would align with our own [292], [294], [295], [296], [297]. Nevertheless, fears that lead to distrust in artificial intelligence are not limited to concerns about AGI. Distrust in AI is attributable based on a variety of factors including (but not limited to) concerns about surveillance, privacy, hackability, autonomous technologies in the defense and transportation sector, and the potential impact of decisions made either directly by or informed by artificial intelligence algorithms [298], [299]. According to one study, the top concerns of those who were distrustful of artificial intelligence were (in order) its uses in war, loss of human control, issues of privacy, applications in healthcare, and potential consequences of artificial intelligence for the economy [299]. Following the concerns about security and defense applications of artificial intelligence, some have argued that we are right to be distrustful of these uses, and that decision making by these algorithms should be heavily circumscribed (requiring human input) [300]. Accordingly, distrust in artificial intelligence tends to increase as the stakes of decision-making increase [301]. Given the high stakes for patients in using artificial intelligence to make diagnoses or suggest treatments, considerable attention has been paid to how to reduce distrust in healthcare settings [223], [225], [302], [303]. However, it should be noted that trust in artificial intelligence in healthcare settings can sometimes outpace trust in human doctors and that this effect is gendered, which raises its own ethical concerns about uses of the technology in a healthcare environment [304]. Finally, there are more general moral concerns about particular applications of artificial intelligence, which roughly map on to concerns about alignment problem for AGI. Here, some worry that, when facing novel circumstances, an artificial intelligence program might exploit vulnerabilities in order to achieve its goals rather than report them [293]. Whereas we might expect (for better or worse) a human competitor to be bound by moral considerations when facing novel circumstances, the response to which is underdetermined by the rules of the competition, a computer may not be so reliable. Whether or not we are right to think that humans are, on average, trustworthy under conditions of competition, it is harder to justify trust in artificial intelligence systems that have not been trained or instructed on what to do under those circumstances.

### 3.3.1.    Distrust makers in AI systems

In what follows, we will elaborate on three major classes of distrust makers in AI systems: surveillance & manipulation, human autonomy & dignity threat, and unpredictable futures. (See, Figure 4)

**Figure 4: three major classes of distrust makers in AI systems**

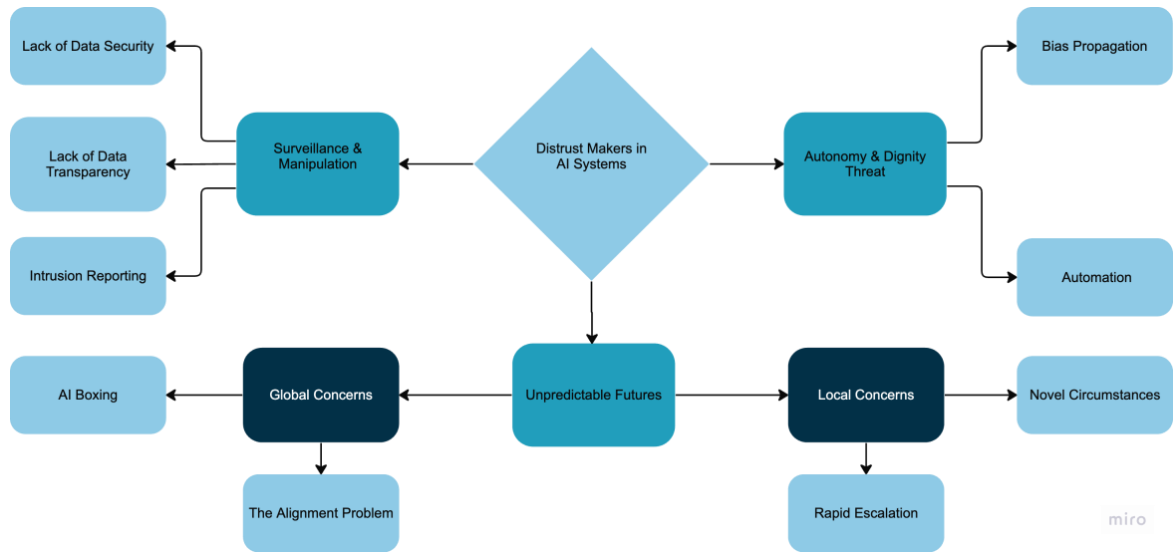

### 3.3.2.    Surveillance, and manipulation

As mentioned above, issues of surveillance and privacy are among the top concerns for those who distrust artificial intelligence [299]. In this case, Applications of artificial intelligence are distrusted because their widespread use might make information about private individuals susceptible to surveillance or data theft [305]. In the case of surveillance, distrust in artificial intelligence carries over from distrust in the parties employing the technology [228], [305], [306]. Distrust of artificial intelligence, in this domain, can range from concern over whether the company or government using the algorithm is trustworthy in what they say about its use to concern over what is said by the algorithm and why it came to that recommendation [228]. These issues suggest that distrust in artificial intelligence is tied to distrust in a particular instance's developers and users. Accordingly, trust in artificial intelligence is a composite of trust in the program itself and in the general scientific and institutional community around artificial intelligence [305]. Further still, distrust in artificial intelligence can also be rooted in distrust of government, even in private applications. This has led some to argue for a layered model in which users first come to trust the government which regulates artificial intelligence developers and then trust the corporate culture, interest, and oversight within the companies which serve as artificial intelligence developers, before coming to trust specific applications or recommendations made by specific algorithms [306]. As for manipulation, distrust in artificial intelligence can be founded in concerns about cybersecurity. For instance, a demonstration by researchers revealed that hacking into the data set of an artificial intelligence program used in a healthcare setting could lead to widespread false detection of cancerous lesions [298]. The potential human cost of systematic misdiagnoses is raised as another contributor to distrust artificial intelligence systems. Concerns like these, have motivated some work on creating artificial intelligence systems, which detect and report outside modification [294]. However, we should also recognize that corrupted data sets are not always the result of outside manipulation, and might merely be incomplete, unbalanced, small, or inaccurate [293].

### 3.3.3. Human autonomy/dignity threat

When it comes to issues of autonomy and dignity, the most prevalent concerns about AI are either 1) that these algorithms will only reify and propagate existing biases and inequities or 2) that artificial intelligence will supplant human agency in part or in total. That is, distrust in AI in these cases is grounded in a skepticism about artificial intelligence's ability to preserve the dignity of all humans and/or human dignity as such. In the first case, considerable attention is and should be paid to the lessons that machine learning (ML) algorithms learn, which might inherit our own societies' biases [227], [303]. As an example, an algorithm designed to predict individual recidivism rates so as to help inform sentencing and parole decisions in the criminal justice system might propagate an existing social bias on the basis of skin tone [293]. If police are more likely to patrol and make arrests in predominantly black neighborhoods, then the data set coded with an increased rate of recidivism will likely follow suit. Thus, even if the artificial intelligence is not specifically looking at race, its data set will have encouraged it to associate facts about race with facts about recidivism. With regards to human dignity in general, and the issue of supplanting human agency, the issue is among the highest rated concerns of those distrustful of AI [299]. Outside of the general concern, distrust in artificial intelligence can be rooted in domain specific intrusions. For instance, despite not always trusting AI, people do sometimes trust artificial intelligence algorithms more than humans (including in healthcare and governance) [304], [307]. The degree to which workers in particular domains find meaning in their work is the degree to which they might perceive the influence of artificial intelligence as pernicious. This effect is amplified when human patients (for instance) come to trust algorithms more than they trust human doctors. The insult to dignity is only made deeper when this imbalance of trust is distributed unevenly along demographic lines (as is the case for female doctors) [304]. Potential solutions to this problem have been proposed, and they often focus on bringing human users into relationship with the artificial intelligence [299], [308].

### 3.3.4.        Distrust and unpredictable futures

Another issue contributing to distrust in AI is unpredictability. Concerns about unpredictability come in global or local varieties. For instance, global concerns include things like those mentioned above, over the large-scale implications for society of machines utilizing artificial generalized intelligence [291], [292], [293]. One species of these worries (again, as mentioned above) concerns the two-pronged project of 1) making sure that an artificial generalized intelligence system has values and interests aligned with those of humans [292], [294], [295], [296], [297] and 2) determining how we could be assured that we succeeded in ensuring this alignment. Likewise, others have suggested that mitigation strategies should be put in place using so-called 'AI boxing' in order to ensure that large-scale social damage is avoided in cases where researchers erroneously believe they have succeeded at both projects [309]. Meanwhile, local concerns include worries about the unpredictability of artificial intelligence systems in specific recommendations or under specific circumstances. One version of this is those, brought up above, of so-called 'dirty tricks', in which novel circumstances are exploited rather than reported [293]. This need not involve circumstances of competition, as without training on how to handle novel cases, machine learning algorithms might simply use personal data in ways that human analysts might not [310], [311]. Another version of these worries includes uses of AI technology in autonomous weapon systems (AWS) [311]. In these cases, AI behavior under novel circumstances (including the so-called 'drone swarming') might lead to conflict escalation too rapidly for humans to intervene so as to avert unsafe or catastrophic outcomes [311].

### 3.3.5.        Challenges and barriers to breaking distrust

There are many challenges and barriers to reducing distrust in artificial intelligence systems. For one, trust in these systems requires some amount of transparency [227], [312]. However, what this norm of transparency entails is less clear. For instance, users and stakeholders are often unmotivated to engage in explanations [226]. In healthcare settings, the utility of these explanations is time sensitive [223]. Explanations of how an artificial intelligence system arrived at a particular recommendation are helpful at critical times but providing explanations at non-critical times tends to diminish trust overall [223]. Likewise, how the explanations are delivered will alter how effective they are at increasing trust. Written narrative explanations are not often as effective as visual representations, and local explanations of specific reasons for recommendations tend to be more effective than global explanations of the artificial intelligence system making the recommendation [223]. Still, there is skepticism about how helpful explanations are in general [224], [225]. As Feldman points out, patients are trusting of healthcare interventions in medical science despite often not understanding the underlying mechanisms at work [225]. We might wonder why we should think that the relative distrust in artificial intelligence is attributable to a lack of explanatory transparency. One challenge is then settling on the right set of conditions for satisfying the optimal level of transparency. Clarifying the conditions required of a good and helpful explanation is a project currently being undertaken in [226], [227], [228]. Likewise, there is a challenge posed by trying to actually assess how trustworthy an algorithm is. Developing models of trustworthiness is an ongoing project [228], [313], [314]. Finally, while there are barriers and challenges to establishing trust in AI, there is also a problem of arriving at the optimal level of trust. People often over-rely on artificial intelligence systems, trusting them too much [226]. For instance, the driver of an autonomous car in Florida crashed into truck because they had over-trusted the artificial intelligence system steering the car [293]. They stopped paying attention to the road and began watching a film during the drive. On top of the challenge of measuring trustworthiness, we then also have the challenge of finding the optimal level of trust and developing interventions, which can push users in that direction as well.

### 3.4.      Trust makers: building/increasing trust in AI

Trust is the essential component for humans to accept AI technology and adopt it in different domains. Hence, technology owners and developers seek strategies to either increase trustworthiness of their AI systems or enhance end-users ' trust. As mentioned in Section 2, trust and trustworthiness are two different phenomena, and one does not necessarily grant the other. Various technical and axiological factors could increase the trustworthiness of AI models, while the literature has paid more attention to technical factors such as explainability and accuracy to enhance trustworthiness. Although trust could be improved as trustworthiness increases, there exist specific trust engineering techniques that only focus on building trust without considering the features of the AI model and its trustworthiness. In this section, we first introduce the overall factors that could affect the trust and trustworthiness of AI to understand the required basis for building trust. We will then review methods of building trust in AI as used in previous research. Finally, we will introduce a few case studies in the different domains and explore their respective influential factors for building trust.

### 3.4.1.    Factors that affect trust

The factors that impact trust in AI systems could be categorized as technical and axiological  factors. From another perspective, these factors could be divided into human-related, AI-related, and context-related factors, where the latter is mostly related to particular requirements of a specific application and the developers' characteristics. Among the technical AI-related factors that influence trust, transparency and explainability have been widely investigated since black-boxes are generally less trustworthy [54].

Among human-related factors, understanding the technology, expertise, culture, and personal traits have been found significant [94]. There are conflicting results about the effect of gender, where it was found effective in [94] but not significant in [315]. However, education and age do not play an important role in building trust. Among AI-related factors, performance and reliability have been significant along with AI personality, anthropomorphism, reputation, and transparency. Finally, team-related factors and risk of the task have been found significant among the context-related factors, where higher risk leads to lower trust [94]. Although several studies have talked about the individual-related factors, findings of one survey study with 226 participants to measure the relative advantage of AI-based advisory over human experts in the context of financial planning suggested that the implementation of human traits was negligible while the ability to test the service noncommittal was superior [316]. While the AI-related factors mostly focus on improving the capabilities of the AI system, the other factors could change trust even when the capabilities of the system and its trustworthiness have not changed. This could lead to over-trust or under-trust, as shown in Figure 5, where the former could cause damages and the latter leads to less adoption of the AI systems [317],[318]. Under trust happens when due to axiological factors such as lack of documentation or good reputation of the developers, the level of trust is lower than the actual capabilities of the AI system.  On the other hand, over-trust happens when the trust is higher than the system's actual capabilities, which could happen due to misperceived trustworthiness. In fact, some of the axiological factors such as reputation, human agency and oversight, and accountability could be engineered true branding, marketing, or other ways. These methods could then lead to over-trust if the AI's ability is not aligned with those factors. Other categorizations of the influential factors have also been proposed in the literature based on subgroups of human-related, AI-related, and context-related factors [319].

**Figure 5: calibrated trust axis** [317]

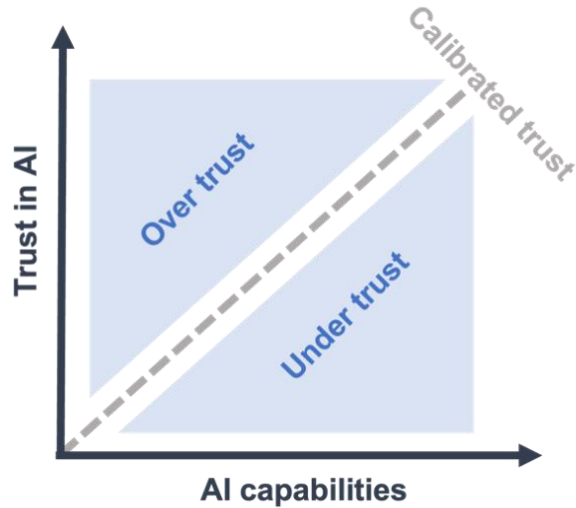

### 3.4.2. Methods of Building trust in AI

Different methods for building trust have been suggested in the literature, where these methods focus on technical factors to enhance trustworthiness or axiological factors that focus on trustors' characteristics to enhance trust [1]. The former methods consider aspects such as model performance, transparency, and explainability. The latter, however, focuses on building trust through accessibility, preparing comprehensive documentation and regulations. From a technical perspective, for an AI system to be trustworthy, technology creators should ensure that the data acquired, processed, and fed into the algorithm is accurate, reliable, consistent, relevant, bias-free, and complete. Similarly, the selected, trained, and tested algorithm should be explainable, interpretable, transparent, bias-free, reliable, and useful [2].

There is an extensive body of literature discussing the explainability and interpretability of AI as one of the most important factors that affect the trustworthiness of AI [2], [18], [85], [320], [321], [322], [323], [324]. Explainability becomes extremely important in critical applications such as healthcare, where the decision made by the AI would not be reliable unless it is justified based on common medical knowledge. It is worth noting that a bad explanation for trust may fail to create trust. In other words, too little detail does not explain trust, and too much detail becomes confusing for users to trust [325].  Another technical method for building trust is to provide confidence level and the AI's decision. When the AI system reports its confidence in its decision, it allows its users to judge how reliable this decision is, and it significantly increases trust [326].

Explainability can be provided in two levels, including global and local explainability. The former explains the overall behavior of an AI model, while the latter explains its decision process in response to a specific input. It was shown that the global explanations about the process had no impact on immediate satisfaction and trust but improved later judgments of understanding about the AI. On the other hand, local justifications were found effective, but their effect is time-sensitive. For instance, during a critical situation or when AI was making errors, local justifications were very effective and powerful explanations [89].

Presentation of the results and explanation of AI systems could also affect the trust. AI systems need to meet a certain level of performance criteria, they need to be explainable and interpretable, they need to consider fairness and biases in their design and evaluation. However, the way that an AI system communicates its results with human agents has a direct effect on trust. A study showed that interactive visualization is a technology that helps to increase trust in AI systems [327]. Another study found that users had significantly more trust in the explanations that were presented by a human agent [328]. In the healthcare domain, it was shown that visual and example-based explanations integrated with rationales had a significantly better impact on patient satisfaction and trust than no explanations or with text-based rationales alone [89].

One of the non-technical methods of building trust through generating and sharing transparent, clear, and comprehensive documentation is the supplier's declaration of conformity (SDoC) [329]. SDoC for AI increases trust by focusing on providing cues to the trustors to understand the system's characteristics better to assess if they will get what they expect from the AI system. As stated in Section 2.1, the availability of accurate and relevant cues is necessary for the trustworthiness of the AI system to be perceived correctly [43]. SDoC is a transparent, standardized, but often not legally required document used to describe the lineage of a product along with the safety and performance testing it has undergone. SDoC gains trust since it shows the process or service conforms to a standard or technical regulation. This document contains sections on performance, safety, and security. It also explains how the system was created, trained, and deployed along with what scenarios it was tested on, how it will respond to non-tested scenarios, guidelines that specify what tasks it should and should not be used for, and any ethical concerns of its use [329]. This level of transparency and detail concerning every aspect of the system, especially the evaluation process, helps increase the trust, but mostly for expert users who know how to interpret the metrics provided in the fact sheet. However, this technique may discriminate against patients from low literacy backgrounds who are less used to interpreting statistical risks [330]. Therefore, in addition to SDoC, there is a need for expert agencies to assess these documents so that non-expert end-users can rely on their assessment. In this case, the experts' endorsement can only function on a principle of value-based trust since this endorsement provides no extra functional information. Trust could be more prominent when this expert agency is the government or a well-known regulatory entity in which people trust. A diverse group of stakeholders could develop and define standards for promoting trust, as well as AI risk-mitigating practices through greater industry self-governance, and adherence to such standards could be verified, specifically through certification/accreditation [331].

Some believe that the public distrust in AI originates from the under-development of a regulatory ecosystem that would guarantee AI's trustworthiness [332]. They argue that being accountable to the public through elaborating rules for AI and developing resources for enforcing these rules is what will ultimately make AI trustworthy enough. Based on this theory, building public trust in AI is not simply a case of creating explainable AI or standardizing various performance metrics for AI components. Instead, public trust requires some authority that urges organizations to take ethical responsibilities seriously and to validate their interpretations of these standards.

Value-based trust also suggests presenting evidence to persuade users that their positive values, for example, inclusion, confidentiality, good-will, are encompassed by developers and governing bodies. This would, in practice, be achieved through the elimination of biases [330]. This could be achieved by requiring either the expert agency or the AI developers to show evidence of previous ethical conduct in data privacy and usage.

To improve trustworthiness, Roszel et al. proposed 20 guidelines that provide clarity on different influential factors, namely, efficacy, reliability, safety, and responsibility of a given AI system [333]. However, it is important to mention that overstating the role of ethics in corporations' policy, culture, and advertisements, known as ethics-washing, in order to avoid or escape governmental regulations and convince and reassure customers to keep with the company's products or services hurts trust [334]. In fact, if users perceive that a company is only pretending to comply with ethical guidelines, they may build less trust in that company. These findings suggest that companies should be aware of the issues of ethics-washing regarding their AI services and should try to avoid ethics-washing in their marketing communication.

### 3.4.3. Case studies and items effects on building trust

Leveraging the aforementioned methods of building trust also depends on the unique requirements and context of different application domains. In the field of marketing, it is crucial to understand how consumer adoption of the information generated by AI can be improved. One of the important aspects of building trust in marketing is understanding the psychological factors that influence consumers' acceptance of AI-generated information and how to induce more favorable consumers' responses about their AI-generated information and marketing. In this domain, the axiological factors could outweigh technical ones such as model performance and explainability. The relationship between number presentation details and users' trust in AI-based marketing was investigated to improve trust as the authors believed that the number preciseness of the recommendation information would critically influence consumer responses to AI technology [335]. The results of this study showed that the use of a precise (vs. imprecise) information format leads to higher trust. Moreover, when the product's objective quality is high (vs. how), information preciseness strongly influences consumers' trust and purchase intentions. Also, interestingly, when the accuracy of the information is low (vs. high), information preciseness has a stronger influence on consumers' responses. This is a perfect example of the importance of customizing trust-building methodology based on the unique requirements of the application domain rather than using a general solution.

Even for the well-known and general solutions of building trust, such as improving transparency and explainability, the implementation of these methods could significantly regulate their impact on trust. For example, in the case of explainability, the impact of virtual agents on the perceived trustworthiness of autonomous intelligent systems was discussed in [336]. It was found that the integration of virtual agents into explainable AI interaction design leads to an increase of trust in the autonomous intelligent system in the particular application of speech recognition. Overall, users had significantly more trust in the explanations that were presented by the agent. The users found the system to be less deceptive, more trustworthy, and less worrying when the explanation results were presented by the agent. This is a great example of using context-based factors to improve trust rather than focusing on the technical aspects of explainability.

Recently, there has been a considerable amount of interest in blockchain technologies. In this area, technical factors of trust and models of trust are important since AI-AI interaction is prevalent in this domain. A platform where consumers and data providers can transact data and/or models and derive value was proposed considering trust complications, given that preserving trust during these transactions is a paramount concern [337]. This study focused on the use of blockchain technology in the field of transfer learning, where a consumer entity wants to acquire a large training set from different private data providers that match a small validation dataset provided by the consumer. Data providers expect a fair value for their contribution, and the consumer also wants to maximize their benefit. To gain consumers' trust, this platform focused on AI-based factors. They implemented a distributed protocol on a blockchain that provides guarantees on privacy and consumer benefit that plays a crucial role in addressing the issue of fair value attribution and privacy in a trustable way.

Some studies tried to understand all the human-related and AI-related requirements of trust in AI-driven chatbots [338]. This study leveraged surveys from end-users and experts and came up with several guidelines and design principles to enhance trust. Another study in the area of human-agent interaction found that human and agents' value similarity plays a significant role in increasing trust. In other words, people base their trust judgments on whether they feel that the system shares similar goals, thoughts, values, and opinions [339]. Finally, to objectively evaluate the factors of trustworthiness of an AI system, a system called Cortex Certifai was developed to assess aspects of robustness, fairness, and interpretability of pre-trained AI models without requiring access to its internal model parameters [340]. Cortex Certifai generates various reports along these axes and only requires query access to the model and an "evaluation" dataset. Using these reports, stakeholders can understand, monitor, and build trust in their AI systems

## 4. DISCUSSION

Despite all the progress in AI, we are in an era of AI similar to when James Watt had to develop the concept of horsepower to help market his steam engine and convince people to buy it. AI is promising a new revolution in technology and has provided excellent results. Nonetheless, AI is complex, and its complexity is an integral part of it. Thus, as James Watt developed the concept of horsepower to gain people's trust in his steam engine, the AI community seeks to adapt trust-building concepts in AI such as explainability, interpretability, and transparency to gain people's trust. Although all of these concepts serve to gain people's trust in AI, it is essential to consider the differences between them and the means to improve each.

### 4.1. Interaction of technical and non-technical factors of trust and trustworthiness in AI

Trust is the essential component in accepting and adopting AI technology in different domains. Although the general definition of trust between humans can be used to define trust in AI, there are unique factors that define trust in AI as a challenging problem. Humans need to ensure that their desires, needs, and rights are fulfilled by the AI; these expectations could be related to AI's performance, reliability, and explainability. It is important to consider the differences between trust and trustworthiness and the means to improve each. While trustworthiness mostly refers to the ability of the AI system and targets technical factors, the trust could be triggered by other non-technical factors such as reputation or documentation. Trust in the domain of AI can be defined in the interaction between human and AI, AI and human, and AI and AI, each of which has some unique requirements beyond the common basic factors of trust. Trust is a critical determinant of the successful adoption of AI technology. A large reason for the lack of adoption of AI models in different domains is the fact that the users are risk-averse and do not implicitly trust AI models. For example, for users to rely on Google Assistant for weather forecasts, they need to trust the information given to them. Lack of trust has significantly limited the application of AI in domains such as healthcare, autonomous vehicles, finance, education, personal assistant, chatbots, etc.

Understanding influential factors of trust is important, but there is an unmet need to understand the relationship between these factors in each domain. In addition to well-known parameters such as performance, explainability, transparency, compliance with certain regulations and standards, and ethical concerns, several other challenges such as bias, discrimination, and privacy need to be addressed to enhance trust and technology adoption further. The most important prerequisite is to identify the factors, their relationship, and how they interact to build trust and make an AI system trustworthy. Doing so requires well-designed experiments and comprehensive models of trust that consider quantitative and qualitative aspects of modeling trust in various domains.

### 4.2. Non-interchangeability of interpretability, explainability and transparency, and their classification

While interpretability sheds light on the relationship between the cause (input) and the effect (output), explainability goes a step further and explains the inside of the AI system and its inner workings. Finally, transparency ensures that these explanations are transparent and clear. Unfortunately, sometimes researchers use these concepts interchangeably. One of the main reasons is the lack of a consensus definition for these concepts in AI. Moreover, it should be noted that different stakeholders need different levels of information. Therefore, the need to classify the degree of transparency, explainability and interpretability for different categories of stakeholders is inevitable. For example, researchers have shown through a behavioral experiment that giving excessive transparency will confuse the user and negatively impact trust [68].

### 4.3. Trust as a two-way street

Another drawback of revealing an excessive level of transparency, which needs more discussion, is that it provides a means for a malicious user to game the system. In other words, trust is a two-way street, so not only must the user gain trust in AI, but AI must also gain trust in the user. Lastly, a famous quote says, "trust is like a piece of paper. Once it is crumpled, it can never be perfect again." Given the presence of giants in the AI industry, such as Google and Facebook, any trust-destructive decisions by big AI companies undermine public trust in artificial intelligence regardless of all academic efforts on XAI. One of the concerns people have when using AI-based solutions is the reliability and safety of AI products. As a result, in addition to academic efforts, the need to establish an institution composed of neutral AI experts without any political, regional, and surveilling biases that oversee the decisions of AI companies and evaluate their products from the perspective of safety and reliability is suggested.

### 4.4. Distinction between empathy in human's trust in AI and empathy in AI's trust in human agents

When it comes to the relation between empathy and trust, a distinction should be made between the role of empathy in people's trust in AI systems and its role in their trust in other persons in computer mediated or online exchanges and communications. Most of the works reviewed above are focused on the former, while some (including [342]) are focused on the latter.

### 4.5. Tradeoff between empathy and privacy.

When empathy is most effective on trust-building, it might involve violations of privacy. This raises issues of tradeoff between the empathy-trust link and privacy. For example, for patients to receive empathic AI-based treatments, they might have to share certain private data. A recent example is Amazon's Alexa's ability to mimic any person's voice,[1] which might enhance empathy and trust. However, such encroachments on privacy might in turn undermine trust. In future work, it might be studied how privacy-breaching empathy might be designed in AI systems (for medical, commercial, educational, and other purposes) so that a circle of mistrust does not ensue.

### 4.6. The subjectivity of trust in AI vs. the objectivity of reliable AI

Trust is a subjective or psychological phenomenon (it is a matter of one's confidence, say, in an AI system), in contrast to reliability, which is an objective probabilistic phenomenon (a matter of whether the system discharges its function properly). This implies that a company might do things (such as creating enjoyment and fun or other presentations), which can attract people's trust, without it being reliable enough. This would result in undue trust or over trust in an AI system, disposing the user to act carelessly with regard to their private information [193]. On the other hand, the subjective character of trust means that a system's reliability does not suffice for attracting people's trust and convincing them to share their private information. Developers of the system need to add features to gain the required degree of trust.

---

[1] Reuters, "Amazon has a plan to make Alexa mimic anyone's voice." URL: reuters.com/technology/amazon-has-plan-make-alexa-mimic-anyones-voice-2022-06-22/.

One might suggest that fairness has a role in enhancing a system's reliability or trustworthiness (as an objective phenomenon), which is a necessary requirement of trust as a psychological state. A system with (almost) no bias is more reliable than a biased system that fails to do justice to all groups of users. Aside from the positive relation between objective fairness and trust, different requirements might be in place for perceive fairness, which is also positively related to trust, including transparency of the data fairly fed into the system or the system's explainability. So, one might explore transparency, explainability, and the like as bridges between objective fairness and perceived fairness.

### 4.7. AI privacy and human agent privacy

There are factors that mediate between trust and privacy, such as explanation of the workings of an AI system or an online provider or transparency, which can engender trust by ensuring that people's privacy is protected. This means that the link between trust and privacy might involve other factors as well, which need to be explored. Moreover, an AI product has private features, the disclosure of which might help gain the user's trust, such as its lineage and provenance. So privacy is a two-way street: both the producer and the user should exchange private data to make the relation work for both.

### 4.8. The developmental problem of 'right to explanation'

It is not enough to guarantee users a 'right to explanation' [343]. Likewise, a right to explanation might create a perverse incentive according to which AI developers are encouraged to produce suboptimal models that are easier to explain in order to avoid the investment necessary to produce an optimal AI system, which is explicable [220]. Another factor leading to this potential perverse incentive is liability concerns on the part of developers [220]. If, as some have suggested, "trustworthiness is… a kind of reliability" then we can distinguish trust in AI from trust in the institutional system that AI emerges from [344]. While these are separate issues and establishing and maintaining the trustworthiness of each requires different kinds of solutions, trust in the latter will increase trust in the former. Likewise, we can further distinguish trust in the larger institutional systems according to whether the concern is about that system will reliably produce AI systems which 1) provide accurate recommendations, 2) provide equitable outcomes, 3) will use data responsibly, and/or 4) will be held accountable for failures of trust.

### 4.9. Development of direct AI accountability

This gestures at a central problem for accountability in AI. For human agents, trust in decision-making and the explanations for our decisions go hand in hand with accountability, but for artificial intelligence systems, decision-making and (potentially) explanation of those decisions rest with the AI, but accountability rests with the developers. For humans, we may trust other humans because we deem their motivations and intentions reliable. At least a part of this is a tendency to act so as to avoid punishment. On straight-forward way to avoid punishment is to avoid punishable behavior. This leads to humans acting cautiously when trusted. It is not clear why accountability for an AI looks like. Yet, without a vision of what it might mean to hold an artificial intelligence system accountable, we have one less tool for establishing the reliability of behavior necessary for trust. In this way, accountability will rest with punishable developers until a theory of direct AI accountability is developed. This will, in turn, engender a perverse incentive for AI developers to avoid liability. Being predictably accurate is often insufficient to establish or warrant trust in humans. Attributions of trustworthiness often require a deeper concern for the basis of this reliability of behavior. This allows us to distinguish a lucky run of correct responses from one brought about because of some reliable mechanism for arriving at correct responses. Since machine learning is often thought of as a 'black box', we may be left only with incentives like punishment avoidance as a potential mechanism for establishing trustworthiness. This leaves a considerable gap in the research when it comes to articulating not just how we might hold researchers and developers accountable for their use and design of AI, but whether it might be possible to hold AI directly accountable (and what this scheme might look like). A robust legal framework will require aligning explanation and accountability at the agential level.

### 4.10. Challenges of measuring trust and trustworthiness in AI

There are many challenges for defining trust and related metrics, including the dynamics of AI, both in terms of moment-by-moment developments and in terms of AI dependence on culture. There are also challenges in measuring trust. Not all of these challenges can be completely solved even with the use of questionnaires, surveys and protocols. If we look at the issue of human selections from a philosophical point of view, we see that the right choice for human beings is also challenging and has different dimensions. It is a person's choices that set him/her apart from the others. Now, how can we define the right choice for implementation in an AI to lead to trust? How would different dimensions (e.g. trust in AI) be concerned without having contradictions in their principles? Ethically, there are similar challenges to evaluate trust. First, without considering different cultures, one must explore what could be the universal ethical principles that can convince everyone to trust in AI. Next, how can these ethical principles be more in tune with a nation's culture in order to achieve greater value in trust assessment? Many cases, by empirical or experimental methods only through trial and error can greatly acquaint us to principles. Psycho-physiological methods that have been used to evaluate trust in human-to-human relationships can also be helpful. Ultimately, all of these approaches should be able to work together with the help of theories to give the correct measures for trust assessment. Evaluating these principles should be used at three levels of organization, group and individual to give a score. These scores can depend on the chosen approach. They include the efficiency of AI and its effectiveness in the task, user understanding, proper interaction between human and AI, control, and data protection. There are other scales to measure, including trust in output and reliance on AI advice, which are also related to efficiency and predictability. Team and individual performance scores, team awareness, and metrics related to this process can also reveal differences between human-human trust and human-AI trust. We must not forget to consider the weakness metrics of the AI system (such as vulnerabilities, errors, and risk assessment) along with the other mentioned metrics. It seems that the ethical framework is the most crucial and serious one (among the various frameworks shown in Figure 3) in codification of principles that can be used for guidelines and protocols by evaluating trust scores. Of course, many frameworks which have been employed in scientific texts are not able to satisfy us that we can work in a singular framework. In fact, it can be imagined that the illustrated frameworks are like nodes of a regular graph that are all interconnected. If the connection of one node to the other on is disconnected, certainly not all aspects of the study will be considered. We know that not all of these can be included in an article with several authors, but they can be defined as a national or international projects for different teams or organizations to find metrics and measure the trust in AI. In this project, the evaluated parameters should be categorized in each framework from the most influential metric to the least important. In this way, a universal comprehensive reference for frameworks and methods of measuring trust in AI via related metrics is designed for all AI manufacturers and its users. This reference includes all principles, protocols, roadmaps, and guidelines for producing and using AI and also trusting them.

### 4.11.    Trust equity problem in AI

In sum, there are a variety of kinds of concerns about AI that result in distrust. While concerns about AGI's have garnered significant academic and popular attention, so-called 'weak AI" is not free from concern. While there is a sizable and growing literature on the reasons contributing to distrust, and on what kinds of explanations count as transparent in a way that encourages trust, there are still many ethical issues raised by considerations of trust in artificial intelligence. First, further research is needed on managing distrust in AI such that automation occurs in equitable ways. Without significant planning and foresight, adoption of AI systems as alternatives to human-centered resources runs the risk of disproportionately affecting human competitors to AI from marginalized groups. For instance, if patients trust artificial intelligence programs more than female doctors but not more than male doctors, then widespread introduction of artificial intelligence could exacerbate professional inequalities in healthcare between men and women. This means that trust in artificial intelligence systems might ultimately determine (in whole or in part) whether automation occurs primarily in industries dominated by otherwise marginalized groups, or (within industries) primarily as a replacement for jobs previously held by marginalized people. Thus, ensuring a basic level of trust necessary for adoption of the technology may not be ethically adequate. Equitable adoption of artificial intelligence entails establishing a robust public sense of trust beyond a minimal threshold. Inversely, further research is needed to determine if the negative impacts of distrust are distributed equitably. If overreliance on artificial intelligence recommendations is domain specific (such that users incorrectly assume that the AI is correct in its recommendation at different levels in different applications or domains), then the externalities associated with this misplaced trust might be distributed inequitably among the stakeholders in that decision process. In this way, concerns like those raised about bias propagation in criminal justice applications of AI might be mediated by judges and lawyers' willingness to grant unearned trust in these systems. This issue, of trust equity in artificial intelligence (which concerns the relationship between trust in AI and equity in AI), demands significant further attention.

The previous discussion highlights the necessity for two significant recommendations for future research. First, researchers should develop and adopt an artificial intelligence trust equity framework. Such a framework would further identify the ways in which trust in artificial intelligence relative to human counterparts is distributed along the lines of demographic data about those human counterparts. This framework would also allow for targeted interventions to appropriately increase or decrease distrust in AI so as to ensure that the effects of artificial intelligence adoption are equitable with regards to economic impact and the concern for human dignity that are wrapped up in automation within the workplace. What a successful targeted intervention looks like is likely to be domain dependent and specific to the particular trust inequities it is designed to target. Second, a complete trust equity framework requires further clarification of the conditions for trustworthiness and for inappropriate trust. This sets up a feedback loop in which solving the ethical issues which arise over equity in AI require research on trust in AI and solving the issue of trust in AI adequately requires research on trust inequity. This suggests the utility of adopting an intersectional approach to analyzing these problems. (See, figure 6)

**Figure 6: AI trust equity feedback loop**

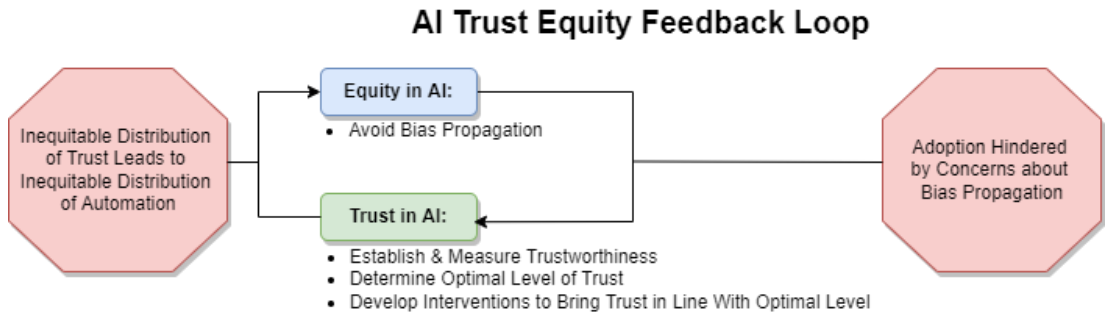

### 4.12. Impossibility of Interpersonal trust in AI systems

The widely accepted assumption is that AI systems cannot have intentions; that is, they cannot intend their functions to be directed at certain goals. On the other hand, a main constituent of human trust is benevolence or honesty, in the sense that we trust in a human agent only when they exhibit honesty and good intentions. For this reason, it can arguably be said that it is not possible to have interpersonal trust in AI systems, since they lack intentions, which is why they cannot exhibit honesty and benevolence, which are necessary components of interpersonal trust. In human interactions, honesty or benevolence provide assurances for a trustful relationship, and absence of this component in the case of human-AI relations makes it more difficult to create trust.

## 5. CONCLUDING REMARKS AND FUTURE DIRECTIONS

Consideration of trust in AI is one of the requirements of developing technologies in the fields of theorizing about AI and designing robots, human-AI interaction, and training their designers and users. In papers under our review, we were able to gain a general graps of factors that would be employed as a metric to work on trust in AI. There are also some basic challenges that must be addressed in future research. To create AI algorithms and products or related technology, in the initial step, we must take the necessary precautions about the care of human and their satisfaction. Moreover, we must be very careful in formulating laws and standardizing AI and related technologies in design and exploiting for all users. These basic principles should be followed by determining the appropriate parameters for product quality remotely or by communication with the user. The implementation of these universal principles is possible only in a pervasive and comprehensive system that can be seen and tracked at any time all over the world. This system must be able to control the growing algorithms and production of technologies, as well as the implementation of principles in their codifications. One of these principles should be assigned to trust definition. This is a challenge that would not be able to achieve just by considering one or more of above-mentioned metrics.

The next step is to consider the dynamics of the AI and technology, which may sometimes conflict with principles written for earlier developments or pre-defined metrics. This may not be a dangerous product in terms of human logic. Therefore, in such cases, the pervasive and comprehensive system must notify all lawmakers of the principles change or modify the previous principles for the new ones. Another example of dynamism is the dependence of AI and technology on different cultures. As we know, different cultures have different protocols, standards, and laws. What one culture deems right may be interpreted as obscenity for another. Therefore, the produced AI should have flexibility within the same boundaries as planned. This flexibility can also be achieved through statistics extracted from questionnaires, interviews, and surveys. Since one of these principles would be related to the concept of trust in AI and its parameterization, we must inevitably work with metrics of trust. Different approaches are employed to give various metrics for trust in AI. A preliminary metric that is necessary to trust in AI is reliability manifested in outputs and proper performance. Furthermore, depending on the purpose of the system, transparency of the system implementation would be in the higher order of significance. The transparency includes explainability, interpretability, and accountability. As mentioned, explainability and interpretability take a higher order of importance than the accuracy. The metric of safety consisting of the fairness, as well as the metric of security consisting of respect for privacy (data protection), are the other factors for trust in AI. The other influential metrics are the provenance (in some texts referred as lineage) and automation of AI. Among these metrics, reducing and/or removing vulnerabilities and errors are very crucial that must be concerned in research. By developing the AI, there would be factors characterized as metrics of trust in the future.

Building trust in AI requires understanding AI-related, human-related, and context-related factors that affect trust in a certain domain. It must be noted that some factors are application-dependent and should be evaluated in the context of the problem at hand. Transparency, explainability, and performance of the AI are amongst the most important technical AI-related factors that play critical roles in building trust in most application domains. These factors mainly increase the trustworthiness of the AI system. However, for the AI system to be trusted by the users, the AI's trustworthiness must be truly perceived by them. This requires certain cues to be provided to the users, which could be done through proper documentation. Other axiological factors for building trust, especially human-related ones, could be engineered to enhance the trust without the need to improve the trustworthiness of AI.

An important need to ensure calibrated trust and avoid over or under trust is to design standards and regulations that could be overseen by trustable agencies such as the government. This approach could increase the trust even among those with less technical knowledge. However, little research has been done into building trust in the growing context of AI-AI interaction. There is an unmet need to design models for calibrating trust in AI-AI interaction as this type of interaction has unique and different requirements compared to human-AI interaction. In the case of AI-AI interaction, parameters such as transparency and explainability have no impact on building trust, while other aspects such as security and reliability become important. In addition, from a technical perspective, there is a need to build robust models against adversarial attacks. Trust models also need to be able to determine malicious resources and calibrate their trust dynamically when interacting with multiple sources.

Finally, after the qualitative literature review, based on the number of reviewed papers and quantitative analysis, we determined that different research eras have not received equal attention. Figure 7 shows what has been done regarding trust-related research in AI, in its four major classes. There are some areas that have received very little or no attention in the literature and may be fertile areas for future research. Some other areas might be open questions for long term.

**Figure 7: Heatmap of the current work distribution on trust semantics, metrics, and measurement in**

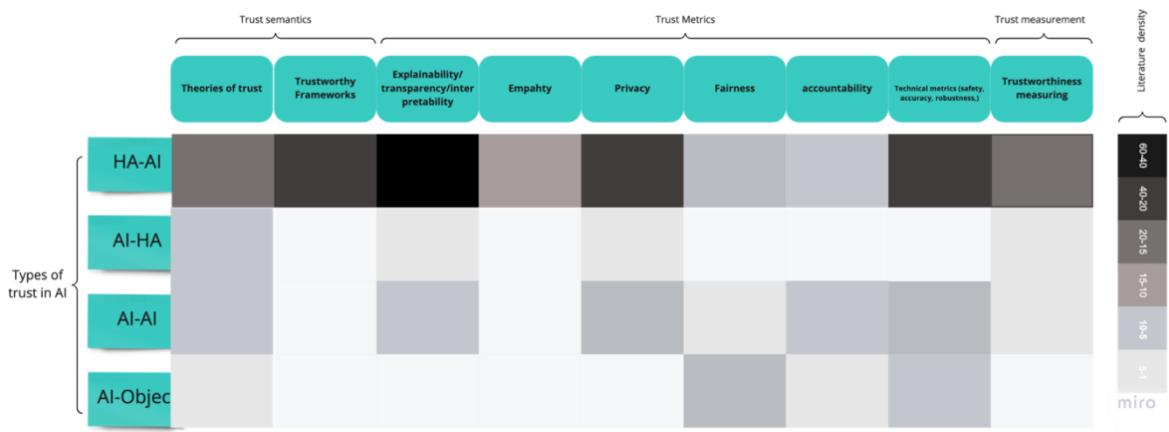

**Conflict of interest**: The authors declare that the research was conducted in the absence of any commercial or financial relationships that could be construed as a potential conflict of interest.

### Acknowledgement

I extend my profound appreciation to Jason D'Cruz and Kush Raj Varshney, with whom I engaged in discussions regarding earlier drafts of this paper during the research process. Their insightful comments and suggestions greatly contributed to the refinement of several sections of the manuscript. Additionally, I wish to express my thanks to Yasser Pouresmail and Mohsen Javaherian for their invaluable feedback and assistance in the initial stages of this study.